\documentclass[useAMS, usenatbib]{mnras}

\usepackage[T1]{fontenc}
\usepackage{ae,aecompl}
\usepackage{graphicx}   
\usepackage{amsmath}    
\usepackage{amssymb}
\usepackage{ulem}

\title[Parabolic jets in high redshift AGN]
{Parabolic jet shape on parsec scales in high redshift AGN}
\author[Nokhrina et al.]{\parbox{\textwidth}{
E.~E.~Nokhrina$^{1}$\thanks{E-mail: nokhrina@phystech.edu}, I.N.~Pashchenko$^{2,1}$,
A.M.~Kutkin$^{3,2}$
}
\vspace{0.4cm}\\
\parbox{\textwidth}{
$^1$Moscow Institute of Physics and Technology, Dolgoprudny, Institutsky per., 9, Moscow region, 141700, Russia\\
$^2$Lebedev Physical Institute, Leninsky prosp.~53, Moscow, 119991, Russia\\
$^3$ASTRON, The Netherlands Institute for Radio Astronomy, Oude Hoogeveensedijk 4, 7991 PD, Dwingeloo, The Netherlands
}
}

\begin{document}

\date{Accepted 2021 October 22. Received 2021 October 22; in original form 2021 June 15.}

\pagerange{\pageref{firstpage}--\pageref{lastpage}} \pubyear{}

\maketitle

\label{firstpage}

\begin{abstract}
Geometry of relativistic jets in active galaxies provides important information about mechanisms of launching, collimation and acceleration of plasma flow.
We propose a new method to probe a boundary shape of a jet on parsec scales -- in the vicinity of its radio core. Apparent speed of an outflow is derived from variability time delays and core shifts measured at the same jet region, providing a self-consistent estimate of the Lorentz factor $\Gamma$. We link together the distance along the jet $z$ with its transverse size assuming a constant flow acceleration. Our results indicate that jets have parabolic shape and sustain an effective acceleration in the core region, consistent with the Lorentz factor dependency $\Gamma\propto z^{0.5}$. The proposed method can be applied to the sources observed at small viewing angles as well as to the distant sources when direct measurements are impossible due to a limited angular resolution.

\end{abstract}


\section{Introduction}
\label{s:intro}

Active galactic nuclei (AGN) phenomenon manifests itself in generating collimated outflows of relativistic plasma -- jets -- and electromagnetic emission with enormous power of $10^{41}-10^{45}\;{\rm erg/s}$. Questions about the nature of AGN activity, role of magnetic field, plasma acceleration and collimation have been addressed for several decades. It is generally accepted that plasma in a jet is accelerated to relativistic velocities by magnetic and electric fields. In contrast with quasi-monopole outflows \citep{BKR-98}, a highly collimated jet is believed to undergo a transformation of initially highly magnetized outflow into a plasma-dominated one \citep{Beskin06, KBVK07, Lyu09}. There is a consensus on a key role of a supermassive black hole rotation in a resulting jet power \citep{McKinney12, BMR-19}. However, many questions remain open. Among them are a jet collimation and a role of an ambient medium. Some of them can be addressed today using data by modern observational programs and experiments. 

The recent studies with Very Long Baseline Interferometry (VLBI) revealed some jets to have a parabolic boundary shape turning into a conical one at hundreds of parsecs downstream \citep{Asada12, Boccardi15_CygnusA, tseng16, r:Nakahara18, Hada18, r:Nakahara19, r:Boccardi_3C264, Kovalev20_r1, r:Nakahara20, r:Boccardi_NGC315, r:Park_NGC315}. Such direct measurements can be obtained for a few nearby sources only due to the limitation of angular resolution achievable with VLBI. It is interesting to probe the shapes of distant sources and check whether they behave in the same way. 

If this phenomenon is common in AGN jets there must be a universal mechanism responsible. Several possible explanations have been proposed to explain changes of a jet shape. In their pioneer work, \citet{Asada12} proposed the change in an ambient medium behavior at the jet shape break. Observed stationary features manifesting standing shocks support this model \citep{Hada18}. \citet{Kovalev20_r1} showed that for seven sources the location of a stationary bright feature coincides with the position of a break of a jet shape. The observed break occurs at $10^5$--$10^6$ gravitational radii $r_{\rm g}$ --- a typical expected distance of a Bondi radius, within which an accretion inflow is dominated by the black hole gravitational potential. This also points to an ambient medium as a determining factor in a jet boundary behavior. On the other hand, a recently detected break in NGC\,315 occurs well within the estimated Bondi radius for this source~\citet{r:Boccardi_NGC315, r:Park_NGC315}. \citet{GL17} suggested a possible model for an abrupt change in a pressure profile due to a disk wind interacting with an ambient medium. \citet{2009ApJ...695..503G} showed that jet can be collimated by a cold wind from the accretion disk and successfully predicted position of recollimation shock in M\,87 jet (HST-1) in their simulations.
\citet{Lyu09}, \citet{BCKN-17} and \citet{Kovalev20_r1} 
proposed that the jet shape break may be present due to a flow being accelerated up to a maximum possible Lorentz factor and transformed to a plasma-dominated regime even for single power-law ambient medium pressure profile. When the energy of electro-magnetic field is transformed into the kinetic energy of plasma the jet magnetization changes. This affects a jet pressure behaviour at the boundary, leading to a different boundary shape for magnetically dominated and plasma dominated regimes.

The jet collimation is closely connected to the bulk acceleration of plasma in the flow~\citep{BMR-19}. \citet{BKR-98} showed that a monopole outflow remains magnetically dominated at any distance from the nozzle. 
\citet{TMN09} showed that axial concentration of magnetic field lines is important for a flow to accelerate. Moreover, only the flows collimated better than parabola accelerate effectively~\citep{KBVK07}. Magneto hydrodynamic (MHD) flow accelerates plasma effectively as $\Gamma\propto r$ up to a half of the maximum Lorentz factor $\Gamma_{\rm max}$. Thus, initially parabolic-shaped flow may be manifestation of acceleration and collimation zone \citep{BMR-19}, while the conical-shaped one represents a jet without any significant bulk acceleration.

Due to limited angular resolution, parabolic jet boundary can be directly observed in the nearby sources. However, there are indications that radio cores of the distant quasars are located in the collimating part of the jet. 
VLBI radio core is the base of a jet seen with VLBI observations. 
This brightest stationary feature in a jet is associated with the
maximum flux at a given frequency or a surface with optical depth equal to unity \citep{BK79}.
\cite{algaba17, r:Algaba-19} found evidence for a non-conical jet shape in the core region using measurements of core size in 57 radio-loud AGNs at several frequencies. \cite{2020MNRAS.499.4515P} showed that after accounting for the bias of the core shift estimation the shapes become consistent with parabolic. \cite{2018A&A...614A..74A} found the cores at 43\,GHz to be more elongated compared to 15\,GHz for a fraction of sources in their complete S5 polar cap sample. This supports a scenario where radio core at 43 GHz resides in a quasi-parabolic jet region.
\citet{kutkin_etal14} found the exponent of the core shift frequency dependence\footnote{Core shift exponent $k_z$ is defined via frequency-dependent position of the VLBI core $z_{\rm core} \propto \nu^{-1/k_{\rm z}}$ \citep{Koenigl81}} $k_{\rm z}$ in range 0.6 -- 0.8 for quasar 3C\,454.3 using different methods. This value agrees with the core of 3C\,454.3 residing in the parabolic flow region, as was shown in simulations of \cite{Porth_etal11}. 
Kravchenko et al. (in prep.) estimated the distance of VLBI cores at 15\,GHz from the jet origin for MOJAVE sample by fitting radial dependence of the jet components size. They estimated the typical $k_z = 0.83\pm0.03$ for MOJAVE sample by comparing the median separation of 15\,GHz VLBI core with the value expected from the median core shift measured between 15.4 and 8.1\,GHz by \citet{pushkarev_etal12}.
Here we propose an implicit method of  
detection of the parabolic jet boundary region for distant sources and sources with small observational angle using the universal connection of geometry and acceleration pattern. The statistical result for a sample of 11 sources points to the quasi-parabolic jet boundary shape in 8$-$15~GHz cores. However, our theoretical prediction in \autoref{Sample} allows us also to explore whether an individual source Lorentz factor at given distance along a jet is consistent with a parabolic domain. 

The paper is organized as follows. In \autoref{Accel} we describe the universal result for a plasma acceleration profile and the connection of a Lorentz factor with a local jet width. In \autoref{Data} we describe the observational data by \citet{kutkin19} (K19 below). We use the full sample to draw a conclusion on a jet boundary shape on the scales of the radio core in \autoref{Sample}. In \autoref{Caveats} we explore the possible caveats, in \autoref{s:coreshift} we discuss possible effects of the results on core-shifts. in \autoref{RandD} We discuss our results in \autoref{RandD} and we summarize them in \autoref{sec:concl}. We adopt the following cosmological parameters: $\Omega_m=0.27$, $\Omega_\Lambda=0.73$ and $H_0=71$~km~s$^{-1}$~Mpc$^{-1}$ \citep{Komatsu09}.

\section{Bulk acceleration} 
\label{Accel}

The Lorentz factor $\Gamma$ dependence on the distance $z$ along a jet may reveal the jet boundary shape profile on the scales up to the observed $z$. This could be done both for the sample of sources from \citet{kutkin19} and for individual sources.
Analytical \citep{Lyu09}, semi-analytical \citep{Beskin06, BCKN-17} and numerical \citep{KBVK07, TMN09}  modelling provides the plasma bulk motion Lorentz factor growth with the cylindrical distance $r$ from a jet axis as
\begin{equation}
\Gamma(r)=\frac{r}{R_{\rm L}}
\label{GammaR}
\end{equation}
for the constant angular velocity $\Omega_{\rm F}$ of field lines. Here the light cylinder radius $R_{\rm L}=c/\Omega_0$, with $\Omega_0$ usually defined as an amplitude  of $\Omega_{\rm F}$. For the model with a constant $\Omega_{\rm F}$ (Model 1, M1) \autoref{GammaR} holds for the full jet width, with the Lorentz factor assuming its maximum value at the boundary \citep{Beskin06, KBVK07, Lyu09}.
For the model with an electric current closed inside a jet (Model 2, M2), $\Omega_{\rm F}$ is constant in the central part of a jet and goes to zero at the boundary \citep{KBVK07, BCKN-17}. In this case, \autoref{GammaR} is valid for this central part. 
Up to now, the linear growth of the plasma Lorentz factor has been obtained only for the magnetically-dominated flow, i.e. while the local magnetization $\sigma$, defined as a ratio of a Poynting flux to the plasma bulk motion kinetic flux, is greater than unity (see Appendix~\ref{a:sigma}). Roughly this condition coincides with $\Gamma=\Gamma_{\rm max}/2$, where $\Gamma_{\rm max}$ is a maximum flow Lorentz factor if a jet electromagnetic flux would be fully transformed into a plasma bulk motion kinetic energy. After reaching $\sigma\approx 1$ the effective acceleration ceases and gradually transits to the logarithmically slow one \citep{BKR-98, TMN09}.

The linear acceleration regime, assumed here, operates i. while the flow is still magnetically dominated; ii. where the field lines curvature is negligible \citep{Beskin06, KBVK07, TMN09, BN09, Nakamura+18}. Indeed, in the magnetically dominated regime the force balance can be written in a force-free approach as Equation (46) in \citet{BN09}. When curvature raduis $R_{\rm c}$ satisfies the condition
\begin{equation}
\frac{R_{\rm c}}{R_{\rm L}}\gg\left(\frac{r}{R_{\rm L}}\right)^3,
\label{Rc}
\end{equation}
the Lorentz factor grows linearly with the radial distance from the axis. When the opposite holds, the pattern of a Lorentz factor growth is different:
$\Gamma\approx\sqrt{R_{\rm c}/r}$, which counterintuitively takes place for a monopole outflow \citep{BKR-98}. For our sources, we have checked that the curvature term in a force balance is indeed negligible. For a field line at the jet boundary set as a parabola
$$z=z_1\left(\frac{r}{kR_{\rm L}}\right)^2,$$
where $k=2.7$ is an empirical coefficient found by K19 and $z_1=1$~pc, the curvature radius is equal to
\begin{equation}
R_{\rm c}=\frac{\left[1+\left(2z_1r/k^2R_{\rm L}^2\right)^2\right]^{3/2}}{2z_1/k^2R_{\rm L}^2}.
\end{equation}
Approximately, the condition of a small curvature is expressed as
\begin{equation}
\frac{R_{\rm c}/R_{\rm L}}{\left(r/R_{\rm L}\right)^3}\approx\frac{4}{k^4}\left(\frac{z_1}{R_{\rm L}}\right)^2\gg 1.
\label{curvcond}
\end{equation}
For the given geometry, while the light cylinder radius is less than 1~pc, we can neglect the curvature term and use the linear acceleration regime (\ref{GammaR}) at the jet boundary. We have checked numerically for our models that the condition (\ref{Rc}) holds not only at the boundary, but also everywhere inside a jet.

This universal acceleration profile allows us to estimate the jet boundary shape having measured the dependence of $\Gamma$ of the projected $z_{\rm proj}$  or deprojected $z$ distance along a jet.
Recent measurements of jet width for a dozen of nearby sources \citep{Kovalev20_r1} provide the transit from quasi-parabolic
\begin{equation}
2r=a_1(z_{\rm proj}+z_0)^{k_1}
\label{rz_par}
\end{equation}
with a power $k_1\approx 0.5$
to quasi-conical shape
\begin{equation}
2r=a_2(z_{\rm proj}+z_1)^{k_2}
\label{rz_con}
\end{equation}
with $k_2\approx 1.0$. Here $z_0$ and $z_1$ are fit parameters, indicating the positions of apexes along a jet.
If the cores are in a region where the acceleration is effective, the observed dependence $\Gamma(z)$ reflects the jet boundary shape:
\begin{equation}
\Gamma=\frac{a_{1,\,2}}{2R_{\rm L}}(z+z_{0,\,1})^{k_{1,\,2}}.
\end{equation}
For the sample as a whole, the acceleration is consistent with parabolic jet boundary shape 
\begin{equation}
\Gamma=(2.7\pm 0.5)z_{\rm core}^{0.52\pm 0.03},
\label{R_Kut}
\end{equation}
where $z_{\rm core}=(R_{8}+R_{15})/2$ is a mean of the observed core positions $R_8$ and $R_{15}$ at the frequencies $8$ and $15$~GHz.
Our aim is to check whether the data for 11 cores from K19 is consistent with the parabolic jet boundary shape.

\section{Observational data} 
\label{Data}

We consider the sample by K19 consisting of 11 radio-loud AGN with measurements of variability time delays using light curves and VLBI core shifts between $\nu_1=15$ and $\nu_2=8$ GHz. Based on these measurements the authors derived the flow speed in the jets and found an evidence for an acceleration of the outflows within a region of an apparent radio core. Namely, a bulk motion Lorentz factor was found to follow $\Gamma\propto z^{0.52\pm0.03}$ on de-projected scales of 0.5--500 parsecs. 
The method to estimate an apparent jet speed based on core shifts and flare delays at different frequencies was first introduced by \citet{KGAA2011}. It is unique in the sense that it allows one to probe the plasma velocity directly in the region of a radio core, unlike the classical approach of measuring apparent speeds of VLBI components. It also may reflect the highest velocity in a jet better than classical kinematics measurements, as the fastest velocities are likely de-boosted.  
The sample includes two radio galaxies, two BL\,Lacertae objects and seven quasars. All the AGN demonstrate extreme variability, which can be explained by maximization of the Doppler factor in a region of radio core (see K19 for details). In fact, the sources were selected to have strong flares in order to measure the corresponding time delays. We underline that though this selection is based on the Doppler boosting, it does not imply any specific intrinsic properties of the jets. Indeed, the maximization of a Doppler factor for a given jet does not imply the maximization of a Lorentz factor in a flow, as the former also depends on the viewing angle. 

The apparent jet velocities measured by K19 are larger than
the ones typically measured by kinematics \citep{MOJAVE_XVII}, but do not contradict them (see discussion in K19). They are
consistent with the velocities of VLBI cores estimated by
\citet{core_shift_var} using the core shift variability data for 40 sources with core shift measurements between 2 and 8 GHz at least at 10 epochs. They modelled core shift variability assuming constant Doppler factor (no acceleration) along the jet and obtained the speed of the photosphere during the flare $\beta_{\rm ph}$.
For five sources that are common between the samples our estimates of $\beta_{\rm core}$ are higher (see Table \ref{tab:compare}). As noted by \citet{core_shift_var}, speed estimates from core shift variability are underestimated because of opacity gradients. So it is not surprising that the limits obtained by \citet{core_shift_var} are lower than our estimates. Another possible reason might be a lower/non-uniform cadence of observational data considered by \cite{core_shift_var}.

\begin{table*}
	\centering
	\caption{Comparing photosphere apparent speed estimates measured by core shift variability from \citet{core_shift_var} and apparent core speed estimates from K19. Columns are as follows: (1) B1950 name; (2) other name; (3) Photosphere apparent speed at 8 GHz, c; (4) Photosphere apparent speed at 2 GHz, c; (5) K19 apparent speed between 8 and 15 GHz, c.}
	\label{tab:compare}
	\begin{tabular}{lllll}
	 Source & Name & $\beta_{\rm ph,8GHz}$ & $\beta_{\rm ph,2GHz}$ & $\beta_{\rm core,8-15GHz}$ \\
	  &  ID & (c) & (c) & (c) \\
	 (1) & (2) & (3) & (4) & (5)\\
	\hline
0607$-$157 &   & 6.3$\pm$4.3 & 11.7$\pm$10.5 & 26.6$\pm$6.0\\
0851$+$202 & OJ 287  & 8.7$\pm$2.4 & 5.8$\pm$1.5 & 30.9$\pm$8.0\\
1308$+$326 &   & 4.8$\pm$1.4 & 3.2$\pm$1.2 & 34.6$\pm$12.6\\
2200$+$420 & BL Lac & 2.1$\pm$0.9 & 3.0$\pm$1.5 & 10.5$\pm$4.8\\
2223$-$052 & 3C 446 & 4.8$\pm$4.4 & 6.0$\pm$2.4 & 40$\pm$11\\

 \hline
 \end{tabular}

\end{table*}

\section{Implicit estimation of jet shapes} \label{Sample}

Four different cases can possibly be realized for the flow at the observed cores. The acceleration regime: the flow can be effectively accelerating (EA) following the relation (\ref{GammaR}) (and transition regime) or the flow is in a saturation regime (SR) with $\Gamma\approx\Gamma_{\rm max}/2$. The jet boundary geometry: core may be either in a parabolic or in a conical domain. Below we probe these possibilities both analytically and semi-analytically.

Within the semi-analytical approach M1 provides the highest Lorentz factor at the jet boundary radius $r$.
In contrast with M1, the model with a closed electric current inside a jet (M2) has a different Lorentz factor distribution across a jet. At a given jet crosscut, it reaches its maximum value $\Gamma_*$ at the field line inside a jet having a radius designated as $r_*$ and goes down to unity at the jet boundary.
Due to differential magnetic field line bunching \citep{KBVK07, TMN09, Nakamura+18}, the radius $r_{*}$ constitute a fraction of a jet radius. Keeping in mind models with non-constant angular velocity (including self-similar), we introduce parameter $\rho\le 1$ as a ratio of a jet radius $r_*$, where the maximum Lorentz factor is reached, to the full jet radius.
Here we discuss the models where $\rho$ is approximately constant for the jet scales under discussion. So,
we use the Lorentz factor dependence of a jet radius in a form
\begin{equation}
\Gamma=\rho\frac{r}{R_{\rm L}}.
\label{Gammarho}
\end{equation} 
Both semi-analytical \citep{BCKN-17} and numerical \citep{KBVK07, Tchekhovskoy19} modelling predict that this fraction $r_*/r$ is roughly $0.2-0.5$ depending on the distance along  the jet (larger $\rho$ corresponds to larger local magnetization). We have found that the empirical relation (\ref{R_Kut}) is an upper envelope for the semi-analytical curves if we set $\rho\approx 0.35-0.55$. This range for $\rho$ is in good agreement with \citep{KBVK07, Tchekhovskoy19}. 
Below for theoretical and semi-analytical estimates we use $\rho=0.5$ for simplicity. In this case the dependence $z(r/R_{\rm L})$ include also the particular value of $\rho$. 

Although we checked all the results for both M1 and M2, below we present in figures only the results obtained either for M1, or for M2. This is because the dependencies, obtained semi-analytically, are the same for reasonably chosen $\rho\approx 0.35-0.55$, and indistinguishable at the figures. The difference of a factor two appears only in particular estimates of physical quantities due to different estimates of $R_{\rm L}$ for M1 with $\rho=1$ and M2. 

\subsection{Theoretical prediction}

The important assumption of K19 work is that all the observed sources indeed represent the same dependence $\Gamma(z)$. Let us show that if we assume the observed cores in the K19 sample are in parabolic and 
EA regime, then all of them must occupy some relatively narrow ``strap'' in $\Gamma-z$ plane. In other words, the assumption made in K19 could be relaxed from the theoretical point of view.

\begin{figure}
\centering
\includegraphics[width=\columnwidth, trim=0cm 1cm 0cm 0cm]{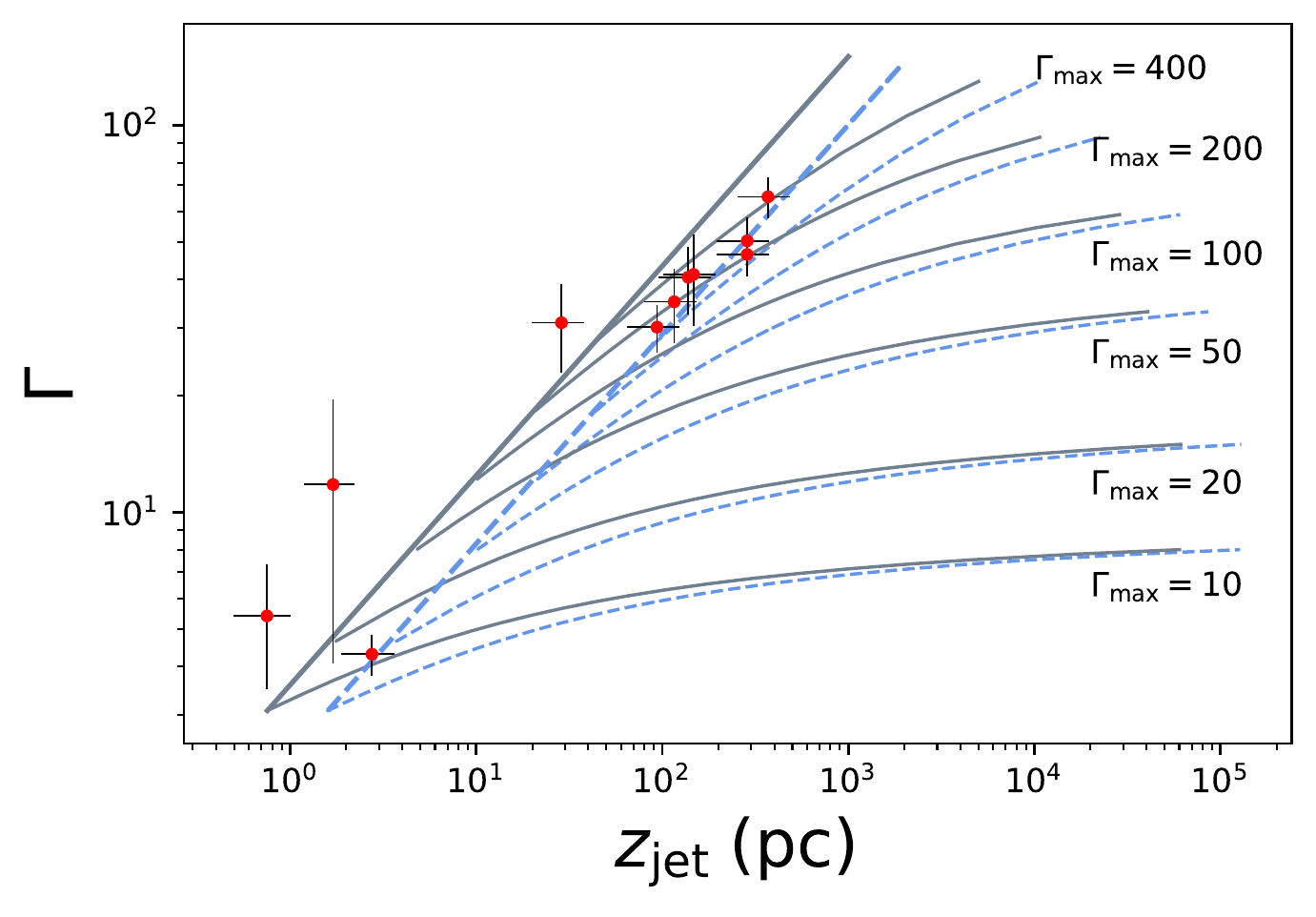}
\caption{Dependence of $\Gamma$ of $z$. 
Semi-analytical modelling (M2) of empirical dependences $\Gamma=2.2z^{0.5}$ (thin dashed blue lines) and 
$\Gamma=3.2z^{0.5}$ (thin solid grey lines) for $\Gamma_{\rm max}=10,\;20,\;50,\;100,\;200,\;400$. Thick dashed blue and solid grey lines conform an envelope of these curves. Red dots are data points from K19 with the errors in black lines.}
\label{envelope_bes_par}
\end{figure}

Suppose that the cores are in a parabolic jet region $r\propto z^{0.5}$, and the jet is in a pressure equilibrium with an ambient medium. Suppose also that the ambient medium follows the pressure dependence of $z$ prescribed approximately by Bondi accretion \citep{QN00, Shch08, NF11}: $P\propto z^{-2}$. In this case we have the conservation of the combination $Pr^4\approx {\rm const}$. 
Using obtained earlier value of this constant  
(see Equation (16) in \citet{NKP20_r2} and discussion therein), we obtain
\begin{equation}
\frac{r/R_{\rm L}}{\sqrt{z}}=
0.63\frac{\sqrt{B_{\rm L}}}{\left(P_0z_0^2\right)^{1/4}},
\label{rz}
\end{equation}
where the characteristic magnetic field is set by the light cylinder radius and the total magnetic flux in a jet: $B_{\rm L}=\Psi_0/\pi R_{\rm L}^2$. Here $B_{\rm L}$ is in the units of G, $P_0$ --- ${\rm dyn/cm^2}$, $z$ and $z_0$ --- pc. We have checked that it holds for both M1 and M2. Numerical coefficient $0.63$ is universal and does not depend on $\sigma_{\rm M}$ \citep{NKP20_r2}. We see that the RHS of this equation depends weakly (as a square root) or very weakly (as a $0.25$ power) on a jet physical parameters: $B_{\rm L}$ and $P_0$ at a fixed distance $z_0$. On the other hand, RHS of \autoref{rz} must be equal to a coefficient $\Gamma/(\rho\sqrt{z})\sim 2.7$ in \autoref{R_Kut} for M1, and $\sim 5.4$ for M2:
\begin{equation}
\frac{\Gamma}{\rho\sqrt{z}}=0.63\left(\frac{B_{\rm L}^2}{P_0z_0^2}\right)^{1/4}.
\label{Gzplane}
\end{equation}
The last equation shows that if flares in different sources with different $\Gamma_{\rm max}$ occur in effective acceleration parabolic domain, their `coordinates' in $\Gamma-z$ plane must lie on one curve, with a coefficient defined by RHS of \autoref{Gzplane}.
Even the large spread in the values of $B_{\rm L}$ and $P_0$, defining this coefficient, leads to the sources being in a narrow band in $\Gamma-z$ plane (\autoref{envelope_bes_par} and Fig.~8 in K19). Here we must emphasize that this does not mean that all the sources in our sample has the same $\Gamma_{\rm max}$. 

The thick blue lines in \autoref{envelope_bes_par} represent the envelope of curves $\Gamma(z)$ for different $\Gamma_{\rm max}$. Here we take the spread in parameters to reproduce the standard deviation of the posterior parameters
distributions for $\Gamma$ in \autoref{R_Kut} obtained in K19. For the envelope we find the corresponding dependence of the jet pressure $P_{10}$ at $z_0=10$~pc distance from the jet base as a function of $B_{\rm L}$:
\begin{equation}
\left(\frac{P_{10}}{\rm dyn/cm^2}\right)=(2\div 8)\times 10^{-7}
\left(\frac{B_{\rm L}}{\rm G}\right)^2.
\label{eq:prediction}
\end{equation}
For reasonable $\Psi\sim 10^{32}\;{\rm G\;cm^{2}}$ and $R_{\rm L}\sim 2-4\times 10^{-3}$~pc, $P_{10}$ is of the order of $10^{-6}-10^{-4}\;{\rm dyn/cm^2}$. The extrapolated pressure estimates for M87 \citep{RF15} and NGC 6251 \citep{Evans-05} down to $10$~pc are $2.2\times 10^{-7}\;{\rm dyn/cm^2}$ and $(4.6-9.2)\times 10^{-7}\;{\rm dyn/cm^2}$, respectively, in a good agreement with our result. 

For an accelerating parabolic jet the coefficient $\alpha_1=a_1\sqrt{\sin\theta}$ in a dependence of a jet width on a de-projected distance $d=\alpha_{1}\sqrt{z}$ can be also related to $\Psi_0$ and $P_0$. Combining \autoref{rz} and \autoref{rz_par}, we obtain
\begin{equation}
\alpha_1=2.3\times 10^{-3}\frac{\sqrt{\Psi_{32}}}{\left(P_0z_0^2\right)^{1/4}},
\label{alpha1}
\end{equation}
where $\Psi_{32}$ is a total magnetic flux in units of $10^{32}\;{\rm G\cdot cm^2}$, and $\alpha_1$ is in units of ${\rm pc}^{1/2}$. We can check the relation \autoref{alpha1} for consistency for M87 and NGC~6251.  
For M87 with $\alpha_1\approx 0.06$ \citep{nokhrina2019} and $P_0=4.5\times 10^{-10}\;{\rm dyn/cm^2}$ at $z_0=220$~pc \citep{RF15}, the estimate for a total magnetic flux is $\Psi_0=1.3\times 10^{33}\;{\rm G\cdot cm^2}$ consistent with \citet{ZCST14, nokhrina2019}. For NGC~6251 $\alpha_1=0.13$ \citep{Kovalev20_r1} and $P_0\approx 3\times 10^{-10}\;{\rm dyn/cm^2}$ at $z_0=480$~pc \citep{Evans-05}. We get the total magnetic flux $\Psi_0=2.7\times 10^{33}\;{\rm G\cdot cm^2}$, consistent with the result by \citet{NKP20_r2}.

\subsection{Effective acceleration regime --- parabolic jet boundary shape} \label{ss:51}

Let us suppose that the observed cores are in the EA
regime.

Within the semi-analytical approach we solve Grad--Shafranov and Bernoulli equations across a jet for a prescribed pressure at the jet boundary \citep{Beskin06, BN09, BCKN-17, Kovalev20_r1}. The natural length unit --- a light cylinder radius, normalizes all the distances in this semi-analytical modelling. Thus, for the M1 we obtain the maximum Lorentz factor across a jet as a function of a non-dimensional local jet radius, at which it is achieved: $\Gamma(r_{\rm jet}/R_{\rm L})$. Combining the empirical relation \autoref{R_Kut} and \autoref{Gammarho}, we set the jet boundary shape as a function $z(r/R_{\rm L})$. After that we are able to calculate every physical parameter of a jet crosscut as a function of a distance along a jet $z$.

\begin{figure}
\centering
\includegraphics[width=\columnwidth, trim=0cm 1cm 0cm 0cm]{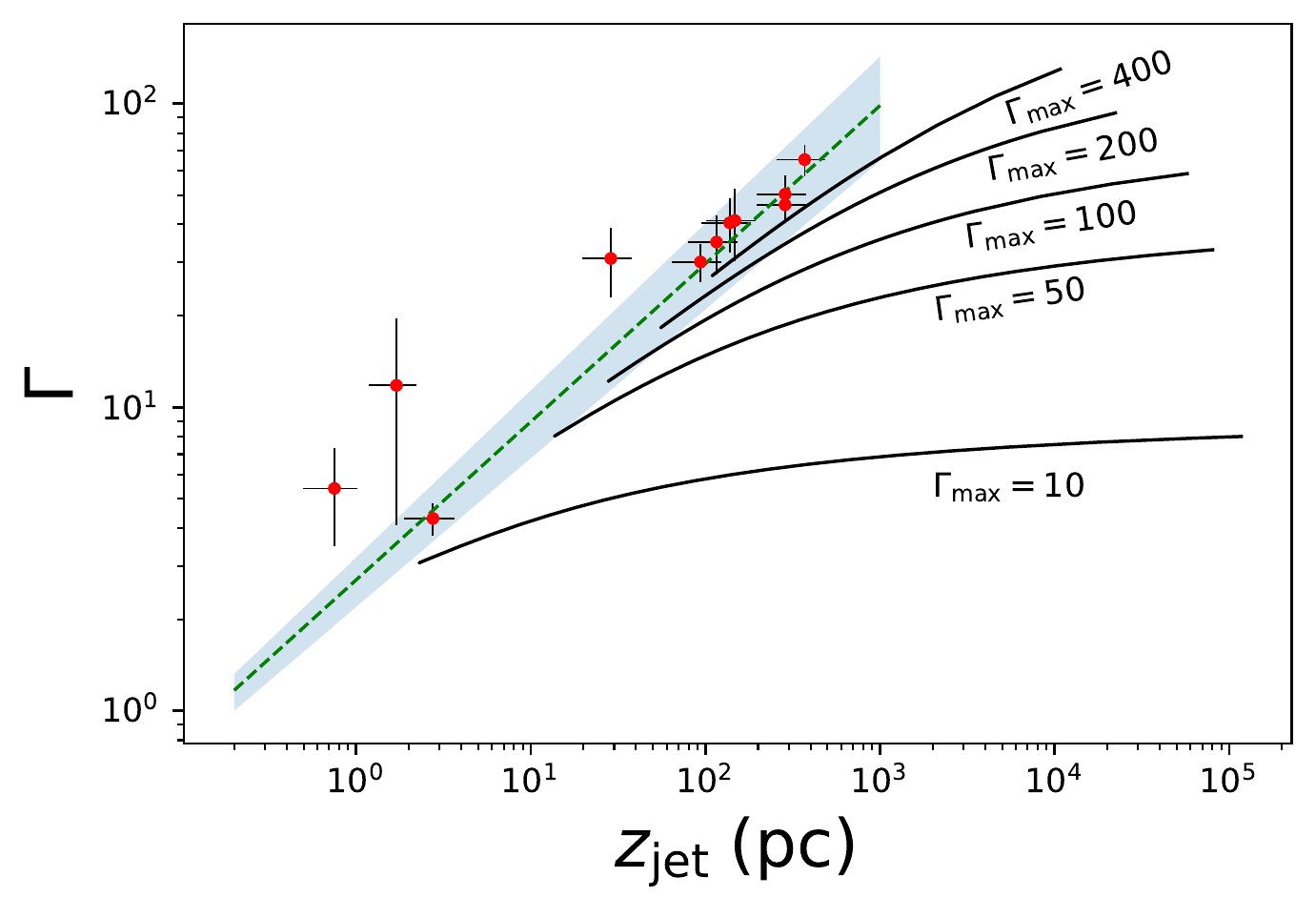}
\caption{Modelling of Lorentz factor for M2 with $\rho=0.5$. Red dots are data points from K19 with errors. Green dashed line is a fit $\Gamma=2.7 z_{\rm core}^{0.52}$. Grey shadowed strip designate errors in the fitting $\Gamma(z)$, from K19. Black solid lines are semi-analytical modelling of a Lorentz factor $\Gamma(r/R_{\rm L})$, where we use \autoref{xofz} to obtain $\Gamma(z)$. Lower value of $\rho$ moves the black lines to the left.}
\label{sample_bes_par}
\end{figure}

For $\rho=0.5$ (M2) the empirical relation \autoref{R_Kut} is an upper envelope of semi-analytical curves $\Gamma(z)$ for different $\Gamma_{\rm max}$ (see \autoref{sample_bes_par}). Thus, the hypothesis of cores being in EA regime in parabolic jet region is consistent with the observational data. The position of data points with regard to the semi-analytical modelling curves provides a rough estimate on the lowest possible $\Gamma_{\rm max}$ for the particular source.
The three leftmost data points may represent a jet with $\Gamma_{\rm max}\gtrsim 10-20$. The seven rightmost points 
correspond to $\Gamma_{\rm max}\gtrsim 100-200$.

Assuming EA regime in parabolic jet, we are able to estimate the black hole spins for the subsample with known BH masses. Combining \autoref{Gammarho} with the \autoref{R_Kut}, we obtain
\begin{equation}
\frac{r}{R_{\rm L}}\approx \frac{2.7}{\rho}\sqrt{z}.
\label{xofz}
\end{equation}
On the other hand, this relation holds at the jet boundary set as $r=(a_1/2)\sqrt{z_{\rm proj}}$,
where we take $k_1\approx 0.5$ for simplicity. 
Approximately we have 
\begin{equation}
r=\frac{\alpha_1}{2}\sqrt{z}.
\label{alph}
\end{equation}
From \autoref{xofz} and \autoref{alph} we relate
a light cylinder radius with a opening coefficient $\alpha$: 
\begin{equation}
R_{\rm L}=\frac{\alpha_1\rho}{5.4}.
\end{equation}
For the  directly measured jet shapes in a dozen of nearby jets the typical value for $\alpha_1$ is $0.068-0.184$ \citep{Boccardi15_CygnusA, r:Nakahara18, Hada18, r:Nakahara19, nokhrina2019, Kovalev20_r1, r:Boccardi_NGC315, Park-19}. This interval may be an upper limit on $\alpha_1$: due to limited linear resolution, we may not detect the parabolic shapes for narrower jets with smaller opening coefficient. Typical $R_{\rm L}$ values for K19 sample and for $\alpha_1$ adopted from the nearby sample is $4.4\times 10^{-3}-1.9\times 10^{-2}$~pc for M2 and about two times larger for M1.

\begin{figure}
\centering
\includegraphics[width=\columnwidth, trim=0cm 1cm 0cm 0cm]{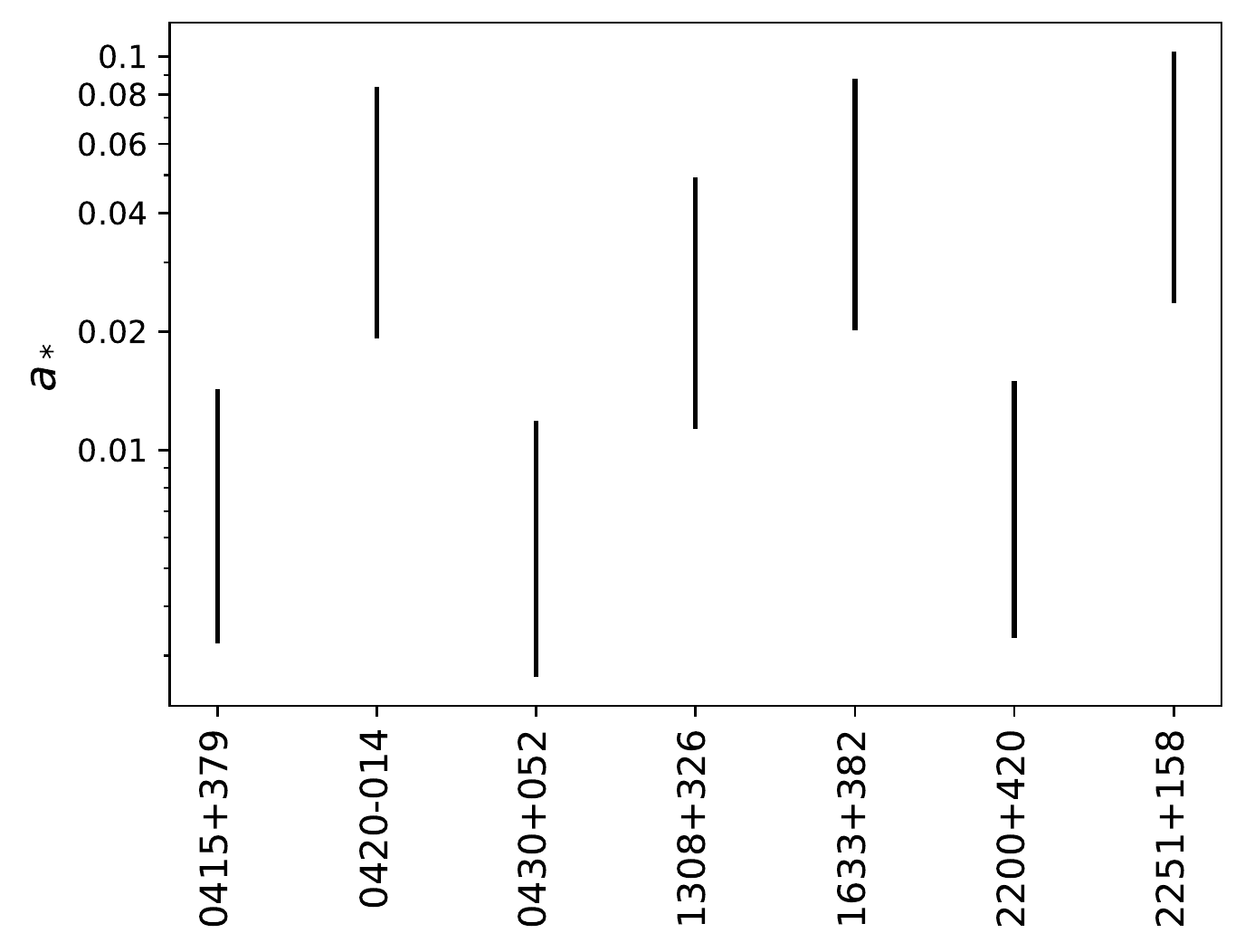}
\caption{Estimates (M2) of $a_*$ for a sample K19 with known BH masses (collected in \citet{Kovalev20_r1}). The dispersion is due to uncertainty in $\alpha_1$ and $\rho$. The upper limit is set by $\alpha_1=0.068$ and $\rho=0.35$. The lower limit corresponds to $\alpha_1=0.184$ and $\rho=0.55$. Values within M1 are about two times smaller.}
\label{f:spins}
\end{figure}

In order to estimate $a_*$ we use the expression for the light cylinder radius as a function of a black hole dimensionless spin $a_*$ assuming maximum jet power condition $\Omega_{\rm F}=\Omega_{\rm H}/2$ \citet{BZ-77}:
\begin{equation}
\frac{r_{\rm g}}{R_{\rm L}}=\frac{1}{4}\frac{a_*}{1+\sqrt{1-a_*^2}}.
\end{equation}
For seven sources in our sample there are mass estimates, collected by \citet{Kovalev20_r1}. The result for spin values is presented in \autoref{f:spins}: we see that spins for 0415$+$379 (3C 111), 0430$-$014 (3C 120) and 2200$+$420 (BL Lac) are less than $0.01$ consistent with \citet{NKP20_r2} (although precise fitting may boost these values up with large uncertainty, see Appenix~\ref{Three}). These sources at least an order of magnitude closer to us than any other source from K19 sample. Interestingly, that this could imply that: i. distant sources with such low spins have too small jet power to be observed (although the jet power depends also on a total magnetic flux $\Psi_0$ and black hole mass); ii. some cosmological evolution of BH spins \citep[see e.g.][]{Barausse12, Sesana14}; iii. some distance-depending systematics in the spin estimation method. The rest of the sources have spin upper limits of the order of $0.1$. However, these values may be a lower limit on actual spins if 
nearby sources indeed provide only an upper limit on $\alpha_1$.
Thus, the robust acceleration profile coupled with the observed relation \autoref{R_Kut} and reasonable estimates of an opening coefficient $\alpha_1$ 
provide us the range of 
black hole spins, marginally consistent with the expected values for active galaxies.

For three sources in our sample (3C 111, 3C 120 and BL Lac) the jet shape break has been detected also by the direct jet boundary shape measurements. We discuss them in Appendix~\ref{Three}. The result of this section for 1633+382 is in agreement with the jet shape measured using the core widths by \citet{r:Algaba-19}.

\subsection{Effective acceleration regime --- conical jet boundary shape}

Let us explore the possibility that a part of our sample (the most distant cores) being at the end of EA
(in a transit from an EA to a SR). The dependence given by \autoref{R_Kut} together with \autoref{GammaR}
rules out the cores being in conical effective acceleration domain. Indeed, the slope of relation between $\Gamma$ and $z_{\rm core}$ for these data points is not a unity. However, at the end of effective acceleration we see the slow bending of $\Gamma(r)$ dependence. Thus, we test the possibility of the farthest seven data points being in this regime at the end of effective acceleration, but still with a strong dependence of $\Gamma$ on $z$.

\begin{figure}
\centering
\includegraphics[width=\columnwidth, trim=0cm 1cm 0cm 0cm]{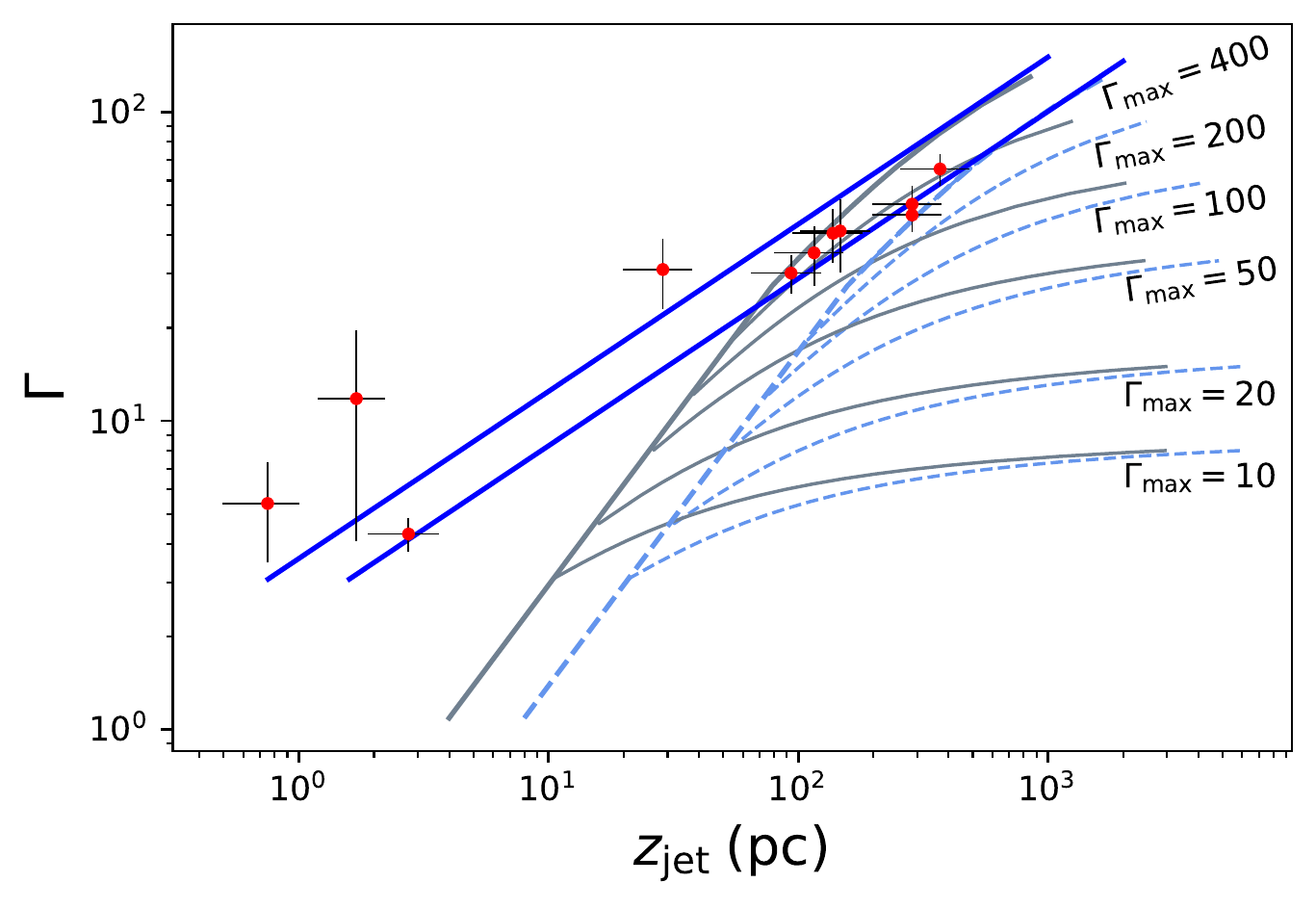}
\caption{Thin lines --- semi-analytical modelling of $\Gamma(z)$ for M2, $c_0=1.25$ (solid grey) and $c_0=2.5$ (dashed blue), $\Gamma_{\rm max}=10,\;20,\;50,\;100,\;200,\;400$. Thick solid grey and dashed blue lines --- an envelope of these curves. Thick blue lines --- envelope for parabolic jet boundary shape from \autoref{envelope_bes_par}.}
\label{Par_vs_Con}
\end{figure}

\begin{figure}
\centering
\includegraphics[width=\columnwidth, trim=0cm 1cm 0cm 0cm]{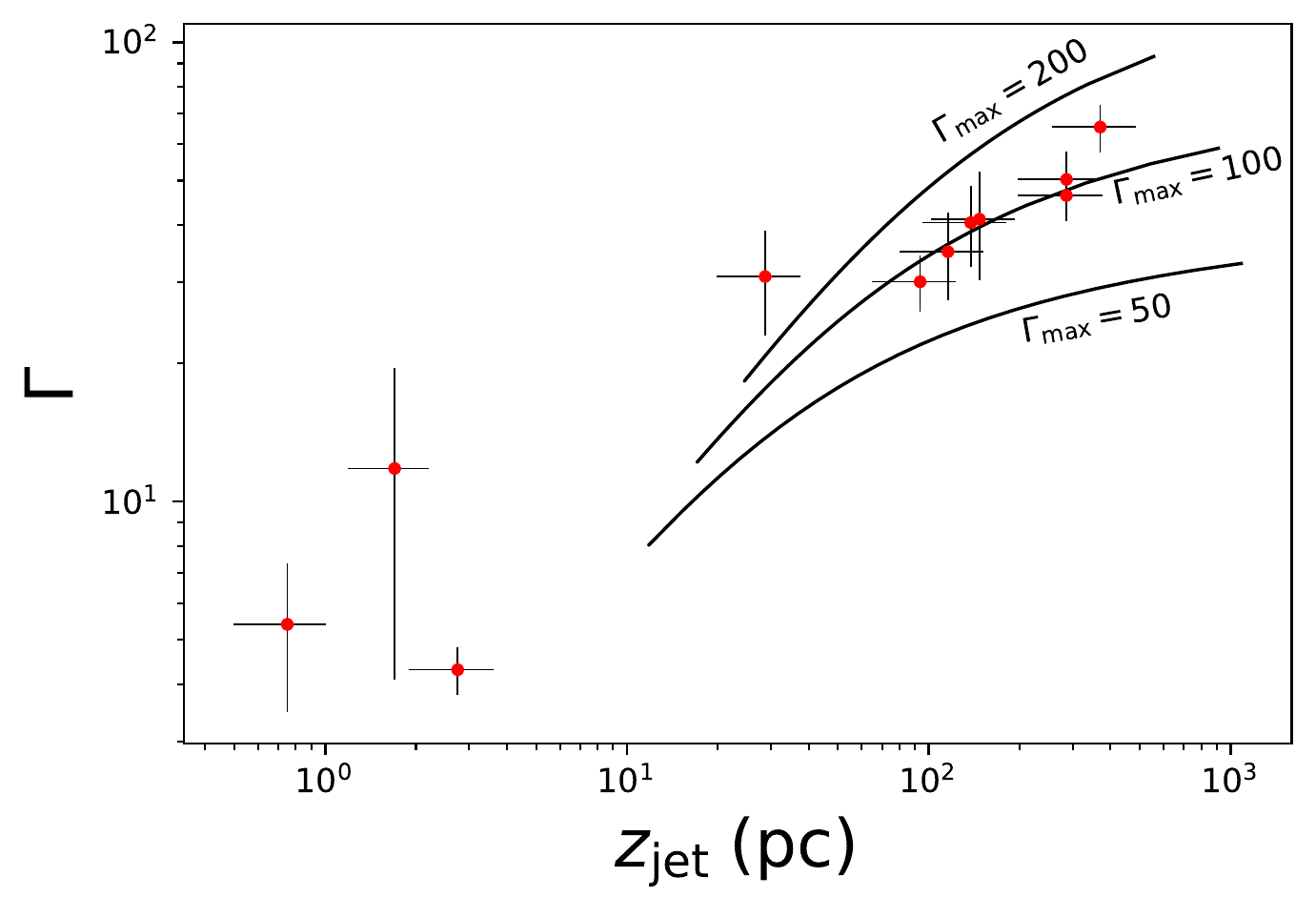}
\caption{Fitting the observational data with conical flow in effective acceleration domain for M1. Here we use $z=0.56 r/R_{\rm L}$, with $z$ measured in pc.}
\label{sample_lyu_con_2}
\end{figure}

To fit the seven furthest data points, we introduce a coefficient \begin{equation}
c_0=R_{\rm L}/\chi
\label{c0}
\end{equation}
to connect distance along a jet $z$ with the jet transversal width $r$ in units of a light cylinder radius: $z=c_0r/R_{\rm L}$. Here $\chi$ is a conical jet intrinsic half-opening angle and $r,\;z,\;R_{\rm L}$ are in pc. This allows us to recalculate semi-analytical dependence $\Gamma(r)$ into $\Gamma(z)$ to fit the data points. 
Coefficient $c_0$ for our data is very well bounded from the theoretical point of view. Indeed, transition to saturation should occur at
transversal jet radius about several times of $r=R_{\rm L}\Gamma_{\rm max}/2$, with $\Gamma\approx\Gamma_{\rm max}/2$ at this point \citep{Beskin06}. Thus, for the sources with $\Gamma\approx 30-65$ the distance of transtition region is roughly $r/R_{\rm L}\approx 100-300$. As $z_{\rm core}\approx 100-350$, the coefficient $c_0\approx 1$. 
We limit the values of $c_0$ using our semi-analytical curves $\Gamma(r)$ for M2.
Because the slope of seven farthest data points is approximately $0.5$, we are not able to fit these points by the relation $\Gamma\propto z$. This limits the coefficient $c_0\approx 2.5$, because for $c_0>2.5$ the data points start to deviate to the left of the curve $\Gamma\propto z$ above even the effective acceleration regime. For $c_0<1.25$ all seven data points correspond to saturation regime with local magnetization $\sigma<1$. So, we limit $c_0\in(1.25,\;2.50)$ for M2. 
The result of semi-analytical modelling for M2 with conical jet boundary shape is presented in \autoref{Par_vs_Con}. Thick green lines stand for envelopes of curves (thin green lines) with $c_0=1.25$ and $c_0=2.5$ chosen to contain the seven farthest data points. 
Values for $c_0$ for M1 are $1/\rho$ times larger. For example,
we have fitted the data with the semi-analytical modelling (M1) for $\Gamma_{\rm max}=200$ with $c_0=1.28\pm 0.11$,
and for $\Gamma_{\rm max}=100$ with $c_0=0.56\pm 0.09$ (see \autoref{sample_lyu_con_2}).

Let us check the consistency of such a fit of the data with a conical geometry and effective acceleration (the transition from EA to SR). We obtained $c_0$ by fitting the data. We estimate the intrinsic opening angles for the sample using apparent opening angles from \citet{MOJAVE_XIV} and observation angles from K19. In this case we can calculate the light cylinder radii using \autoref{c0}. The values of $R_{\rm L}$ needed to reconcile data points with conical geometry are in an inteval $R_{\rm L}\in(0.0016;\;0.025)$ parsec. For the typical black hole mass $M=10^9\;M_{\odot }$ the left interval end corresponds to $a_*=0.25$ while the right one corresponds to $a_*=0.016$. Although the latter value is not expected in powerful jets, it is consistent with the results by \citet{NKP20_r2}. 

Although the half-opening angles and typical $R_{\rm L}$ are consistent with the far cores being in a conical jet boundary shape and an effective acceleration domain, we think that this scenario is unlikely.
In \autoref{Par_vs_Con} the envelope of dependence $\Gamma(z)$ falls very steeply. The bulk flow Lorentz factor becomes equal to unity at distances of the order of 4$-$8~pc. This, in turn, means that the jet should retain the width of the order of a few of light cylinder radii at these distances. The external pressure supporting such a thin jet should be a few of $10^{-3}\;{\rm dyn/cm^2}$ up to $10^6\;{\rm dyn/cm^2}$ for $\Psi=10^{32}-10^{34}\;{\rm G\,cm^2}$ --- several orders larger than extrapolated inside from the measured values \citep{RF15, Evans-05}. 

The scenario of seven farthest cores having a break from a parabolic to conical boundary shape seems artificial. Indeed, the ``strap'' is theoretically expected for every core in the parabolic regime. Thus, if seven farthest cores were in the parabolic regime and switched to the conical somewhere, they should lie above the strap. On the other hand, let us assume these seven sources indeed have a break --- a transition from parabolic to conical. In this case, their ``track'' in $\Gamma-z$ plane should be at first parallel to blue lines but below them (see \autoref{Par_vs_Con}), then get off the track at each individual point and follow conical trend. Finally, they have to get to the initially parabolic strap between two blue lines. It seems unlikely that the sources with a complicated individual tracks with a break should end up in one narrow ``strap''.

\subsection{Saturation regime --- conical and parabolic jet boundary shape}
\label{ss:44}

\begin{figure}
\centering
\includegraphics[width=\columnwidth, trim=0cm 1cm 0cm 0cm]{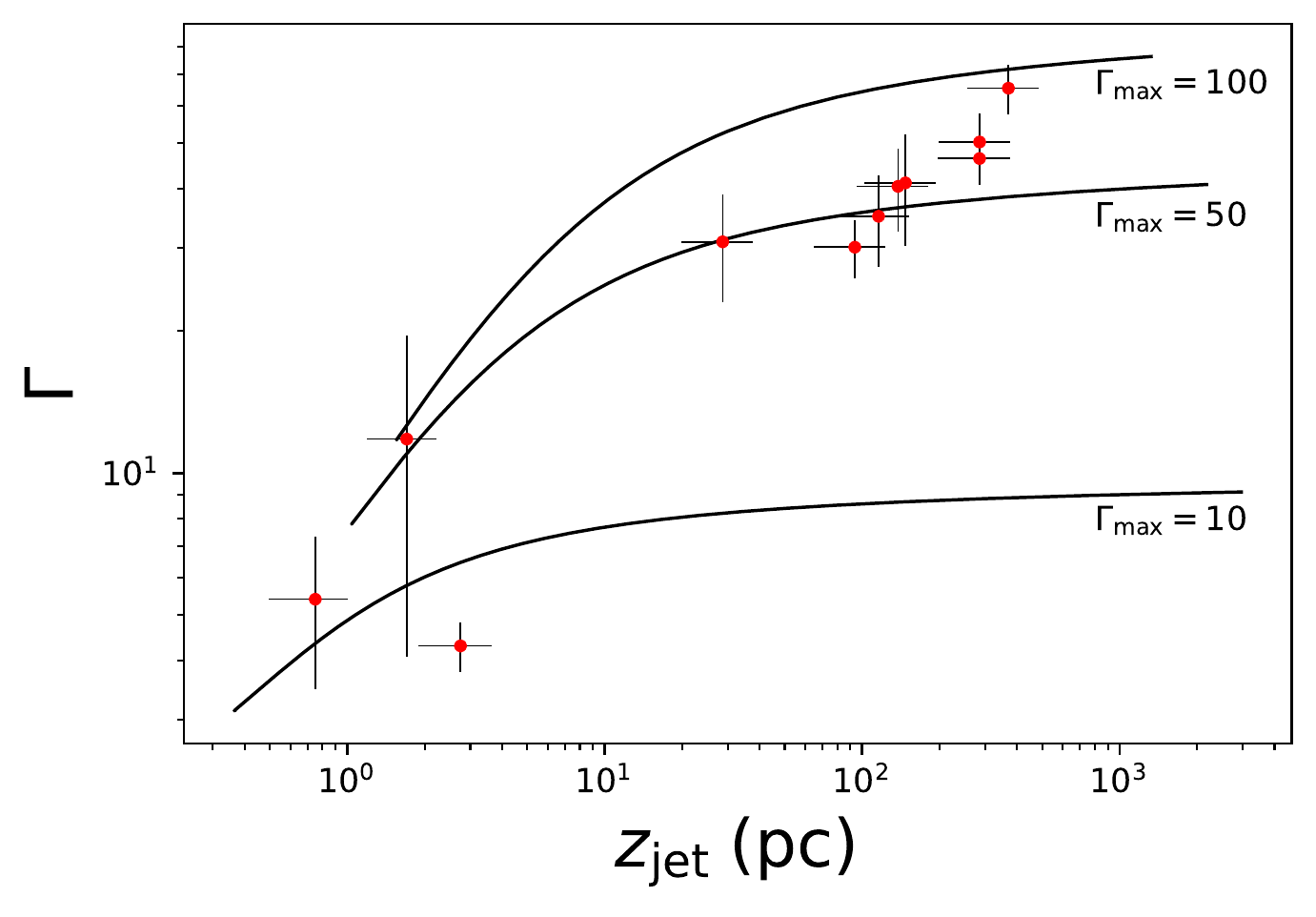}
\caption{Modelling of Lorentz factor with M1 for a conical flow. We set the jet intrinsic opening angle as $\chi=2^{\circ}$ and $R_{\rm L}=4\times 10^{-3}$, which corresponds to $M_{\rm BH}=10^{9}\,M_{\odot}$ and $a_*=0.1$.}
\label{sample_lyu_con_1}
\end{figure}

Setting reasonable typical light cylinder radius (see \autoref{ss:51}) and a jet half-opening angle, we have $\chi\rho/R_{\rm L}\approx 5-175$. This value puts the effective part of $\Gamma(z)$ to the left of the observational data (see \autoref{sample_lyu_con_1}), meaning the cores are in saturation regime.

Modelling jets kinematics for conical jets (see \autoref{sample_lyu_con_1}) corresponds to all the sources except for the slowest three are in a saturation regime with maximum Lorentz factors $\sim 50-100$, which is in agreement with \citet{MOJAVE_XVII}. However, in this case every source is on its own acceleration track, which contradicts to the obtained by K19 correlation. Indeed, the value of $\Gamma$ for each source is approximately equal to $\Gamma_{\rm max}$ and should depend on $z$ logarithmically weak. This would mean that the correlation \autoref{R_Kut} is accidental. The same holds for M2.

Thus, we think that the assumption of a conical jet geometry is unlikely for the available data for both conical and saturation regime. The same holds for a parabolic boundary shape and saturation of acceleration.

In subsections \ref{ss:51}--\ref{ss:44} we tested four possible scenarios for the observed cores. 
\begin{itemize}
\item The assumption of the cores being in EA regime, parabolic boundary shape, is consistent with the data points. We are able to estimate black hole spins for the subsample with known masses. Four spins have an upper estimate of the order of $0.1$, marginally consistent with the expected spin values for radio loud AGNs. Three sources have spins of the order of $0.01$ far from the expected values, but consistent with estimates based on a measured jet width at the break point \citep{Kovalev20_r1, NKP20_r2}.
\item The full sample is not consistent with EA conical jet shape scenario, so we tested it for seven farthest cores. Although this data subset is in agreement with cores being at the end of EA conical regime, and estimated spin values are reasonable, the pressure needed to support such a narrow jet is several orders higher than typically measured \citep{Evans-05, RF15}. Thus, we think this scenario is unlikely.
\item If we assume the cores are in SR (parabolic or conical), we would not expect any correlation of the type obtained by \citet{kutkin19}. So we think this scenario is not consistent with the data.  
\end{itemize}

We conclude that 
the observed cores are in a region of a jet having a parabolic boundary shape and plasma in EA regime --- the so-called acceleration and collimation zone (ACZ, \citet{BMR-19}). This finding supports the previous result by \citep{NKP20_r2}. The fitted semi-analytical acceleration profiles suggest that the initial magnetization may be high for these sources --- up to several hundreds. In this case we should expect high velocities in some jets, which may not be observed due to de-boosting. On the other hand, the cores with highest Lorentz factors may lie at the very end of effective acceleration. In this case the observed velocities of the order of several tens are close to their maximum values.  

\section{Caveats} \label{Caveats}

There are caveats that might affect the obtained result. The first issue may originate from the actual location of the highest Lorentz factor across a jet. Suppose there is a jet with field lines bunching in such a way, that while the boundary (or observed boundary) line retains one shape $r\propto z^{k_1}$, while the line with the highest Lorentz factor retains a different shape $r\propto z^{k_2}$. In this case, the presented above analysis will display the second shape, not the actual behaviour of a boundary. However, we see from the semi-analytical modelling by \citet{BCKN-17} and by numerical modelling by \citet{Tchekhovskoy19} that the position of the highest Lorentz factor occurs at more or less the same fraction of a jet radius, with very slow decline inside. In this case our results hold. The interesting test of this effect could be performed if in some sources from our sample the conical jet shape is discovered by direct measurements. This would imply that jet shapes observed in intensity might consist of effectively accelerating parabolic central part, giving its own imprint in acceleration along a jet, enclosed in some bright slower sheath flow with different geometry. There are some indications of this in \citet{algaba17} (sources 0607$-$157, 1730$-$130 and 2223$-$052), although with low goodness of a fit.

\citet{2020MNRAS.499.4515P} used simulations with the relativistic jet model of \citet{BK79} and found that core shift values obtained by a Gaussians modelling are typically overestimated by a factor 1.5. This could affect both speed and distance estimates. However, as core shifts are measured generally during the quiescent state and they increase during the flare \citep{core_shift_var} at nearly the same scale, these two systematic effects should compensate each other.

K19 estimated the distance of the core from the jet origin $z \propto \nu^{-1/k_z}$, where $k_z$ -- was taken to be 1. In the collimating jet it could be different ($k_z < 1$), hence providing smaller $z$. However, the assumption of $k_z=1$, that holds for a conical jet in equipartition with $B\propto r^{-m}$ and $n\propto r^{-n}$, $n=2m$, does not affect our results in considering parabolic jet boundary at the observed cores. Indeed, the only measured distance is a core shift $\Delta z=z_{8}-z_{15}$. If we assume the conical jet with $k_z=1$, the mean position of cores $z_{\rm core}=(z_8+z_{15})/2$ is situated at 
$$z_{\rm core}=\frac{\nu_1+\nu_2}{2(\nu_1-\nu_2)}\Delta z\approx 1.64\Delta z$$ 
from the jet base, corresponding to the infinite frequency. We take this distance as a proxy to the mean core position. If we assume the strictly parabolic jet shape with $k_z=0.5$, then the distance of the core from the jet base is smaller and is equal to 
$$z_{\rm core,\,par}=\frac{\nu_1^2+\nu_2^2}{2(\nu_1^2-\nu_2^2)}\Delta z\approx 0.90\Delta z.$$ 
If we use this, we have the following change in a coefficient without any change in a power for the empirical relation between $\Gamma$ and $z_{\rm core}$: $\Gamma=3.7z_{\rm core,\,par}^{0.52}$. This will not affect our results qualitatively. However such numerical estimates as black hole spin will become larger, so the results presented in \autoref{f:spins} should be treated as a lower boundary.
    
However, we do not expect such high deviations of $k_z$ from its canonical value. Indeed, \citet{kutkin_etal14} found $k_z$ in range 0.6 -- 0.8 for quasar 3C454.3. This agrees with our finding that the core of 3C454.3 resides in the accelerating parabolic flow. \citet{Porth_etal11} found $k_z \approx 0.9 - 1.0$ for their simulated multifrequency synchrotron maps of RMHD with equipartition between the magnetic and emitting particles energy densities. They attributed $k_z < 1$ to the jet collimation.

Another caveat is the assumption of the constant flow velocity made in K19 to estimate the bulk flow motion Lorentz factor. Below we explore the possibility that the flow between cores at 15 and 8 GHz is still effectively accelerating, which is in obvious contradiction with the assumptions made in calculating $\Gamma$. Thus, we should show that our use of the results by K19 is justified. In Appendix~\ref{a:velocity} we assume the flow is accelerating at the rate $\Gamma_i(z)=a_i\sqrt{z}$ for each source designated by the index ``$i$''. We calculate the relation between the distance $\Delta z$ along the jet in the source frame and the observational time $\Delta t_{\rm obs}$ for such an accelerating flow. After that we pick such a coefficient $a_i$, that the flow  with the constant velocity $\beta_{\rm const}$ (which corresponds to $\Gamma$) crosses $\Delta z$ at the time $\Delta t_{\rm obs}$. After defining $a_i$, we calculate the actual Lorentz factor at $z_{\rm core}$ of an accelerating flow. It turns out that the difference is negligible (see Appendix~\ref{a:velocity}).

\section{Core-shift biases} \label{s:coreshift}

Location of cores in a parabolic jet region leads to possible core shift biases.
Assuming collimation profile $r \propto z^{k}$, radial profiles of the Doppler factor $D \propto z^{-\delta}$, the magnetic field $B \propto z^{-m}$ and the emitting particles density amplitude $N \propto z^{-n}$ (both in a plasma frame) and making the same approximation as \citet{BK79}, the exponent of the core shift frequency dependence:
\begin{equation}
    k_z = \frac{(\delta+m)(1.5+\alpha) + n - k}{2.5+\alpha}
    \label{k_distance}
\end{equation}
where the optically thin spectral index $\alpha$ is defined as the flux density $S_{\nu} \propto \nu^{-\alpha}$. 
This relation can be re-written in terms of the exponents of the radius dependence of the quantities, i.e. $B \propto r^{-b_{\rm r}}$, where $m = k b_{\rm r}$:
\begin{equation}
    k_z = k\frac{(\delta_{\rm r}+b_{\rm r})(1.5+\alpha) + n_{\rm r} - 1}{2.5+\alpha}
    \label{k_radius}
\end{equation}
For $\alpha = 0.5$, constant Doppler factor $\delta_r = 0$, toroidal dominated magnetic field $b_r = 1$ and $n_r = 2$ the core shift exponent $k_z = k$.

However, the non-conical jet shape does not affect core size dependence on frequency. Indeed, as $\theta_{\rm core} \propto z^k$, substituting $z \propto \nu^{-1/k_z}$ with $k_z$ from (\ref{k_radius}), we obtain $\theta_{\rm core} \propto \nu^{-1/k_{\theta}}$, where $k_{\theta}$:
\begin{equation}
    k_{\theta} = \frac{(\delta_r+b_r)(1.5+\alpha) + n_r - 1}{2.5+\alpha}
    \label{k_theta}
\end{equation}
does not depend on the geometry and $k_{\theta} = 1$ for canonical $b_r = 1$, $n_r = 2$, $\alpha = 0.5$ and constant Doppler factor $b_r = 0$.

For the accelerating flow we must take into consideration the magnetic field and the particle number density in a nucleus (lab) frame $N_{\rm lab}=N\Gamma$ and for dominating toroidal magnetic field $B_{\rm lab}=B\Gamma$. For $\Gamma\propto z^{k}$ we obtain
\begin{equation}
\begin{split}
    k_{\rm z}^{\rm acc} = \frac{(\delta+m+k)(1.5+\alpha) + n }{2.5+\alpha}\\ = k\frac{(\delta_r+b_r+1)(1.5+\alpha) + n_r }{2.5+\alpha}
    \label{k_distance_acc}
\end{split}
\end{equation}
In the accelerating jet with LOS angle $\theta$, the Doppler factor increases along the jet up to the maximal value $D_{\rm max} \approx 1/\sin\theta$ at the distance where the bulk Lorentz factor $\Gamma \approx D_{\rm max}$ (K19). Then $D$ decreases up to the region where the jet acceleration stops and remains constant further out. However, if the maximal attainable $\Gamma < 1/\sin\theta$, Doppler factor never decreases along the jet.
Assuming Doppler factor $D \propto \Gamma$ (i.e. $\delta_r = -1$) for jet region with $\Gamma < 1/\theta$, standard scalings for the toroidal magnetic field and particles number density, we obtain $k_{z}^{\rm acc} = 4k/3$ and $k_{\theta}^{\rm acc} = k_{z}^{\rm acc}/k = 4/3$.
\citet{r:Park_NGC315} obtained $k = 0.58\pm0.05$ for NGC~315 while they estimated $k_z = 0.72\pm0.11$, that is close to the predicted $k_z = 4/3k = 0.77\pm0.07$.

Kravchenko et al. (in prep.) estimated the typical $k_z = 0.83\pm0.03$ for MOJAVE sample by comparing the median separation of 15\,GHz VLBI core with the value expected from the median core shift measured between 15.4 and 8.1\,GHz by \citet{pushkarev_etal12}. Assuming on-going bulk acceleration in this region, we obtain the typical geometry profile exponent $k = 3k_{z}/4 = 0.62\pm0.02$ for the MOJAVE sample, consistent with the proposed acceleration and collimation in the core region.

Using data from \citet{puskov15} between 2 and 8 GHz for a non-scattered subsample of $\approx$3000 AGN that lie away from the Galactic plane, we obtained\footnote{Note that \citet{puskov15} uses the inverse definition of $k_{\theta}$: $\theta_{\rm core} \propto \nu^{-k_{\theta}}$.} the median $k_{\theta} = 1.069\pm0.014$ that significantly deviates from zero. This is even more pronounced for a subsample of 342 sources with more precise $k_{\theta}$ estimates obtained using at least four frequencies between 2 and 43 GHz with a median $k_{\theta} = 1.133\pm0.014$.
Interesting, that comparing $k_{\theta}$ at low and high frequency triplets for 76 sources with core size measured at all six frequencies reveals higher median values at higher frequency range (15, 22 and 43 GHz) frequencies: $k_{\theta}^{\rm high} = 1.43\pm0.10$ and $k_{\theta}^{\rm low} = 1.11\pm0.04$ for 2, 5 and 8 GHz. This result is confirmed by the Anderson-Darling test which showed a significant ($p$-value $< 0.001$) difference in the $k_{\theta}$ distributions.
Thus, data from \citet{puskov15} could imply that radio cores at least at frequencies higher than 15 GHz are located in the acceleration parabolic domain.

However, \citet{puskov15} used simple two circular Gaussian model to fit the source structure. \citet{puskov12} used a much smaller sample of 370 sources, but more detailed Gaussian models of their structure and an elliptical Gaussian model for the core at 2 and 8 GHz. We estimated the median $k_{\theta}$ using their data for all sources $k_{\theta} = 0.99\pm0.03$ and BL Lac and quasars only: $k_{\theta}^{\rm bl} = 0.95\pm0.08$ and $k_{\theta}^{\rm qso} = 1.00\pm0.04$. Although the obtained values are consistent with $k_{\theta} = 1$, they are significantly different from the value based on \citet{puskov15} multifrequency data set. This could imply that core size measurements are biased if done with a simple (e.g. two circular Gaussians) models or if the core is fitted with a circular Gaussian. If it is indeed the case, then data of \citet{puskov12} implies that radio cores at frequencies lower than 8 GHz are located in a conical constant speed domain\footnote{Strictly speaking, $k_{\theta} = 1$ holds also for the accelerating jet with $\delta_r = -1.5$ \autoref{k_distance_acc}. However, we expect a slower growth of a Doppler factor in $\Gamma < 1/\theta$ region.}.

Expression \autoref{k_radius} for conical jets leads to a classical expression found by \citet{Koenigl81}, and for a parabolic boundary shape it gives $z_{\rm core}\propto\nu^{-2}$. Thus, if we measure core positions at two frequencies $\nu_{\rm high}$, $\nu_{\rm low}$ and estimate the jet base position assuming conical relation $z_{\rm core}\propto \nu^{-1}$, we probably overestimate the distance to a physical jet base. Indeed, the distance to the jet origin from the core at $\nu_{\rm high}$:
\begin{equation}
    z_{\rm high} = \frac{\Delta z}{ \displaystyle\left(\frac{\nu_{\rm high}}{\nu_{\rm low}}\right)^{1/k_z} - 1}
    \label{eq:zhigh}
\end{equation}
where $\Delta z$ -- core shit between two frequencies. In a collimating jet with a gradually increasing $k$ \citep[e.g. in MHD model of][]{BCKN-17} this systematics remains even if one employs the measured $k_z$ instead of $k_z = 1$ for estimating the distance to the jet origin. Thus, to infer the jet geometry from the core shift effect, one should better use the ratio $k_z/k_{\theta}$ to obtain the shape exponent $k$ corresponding to a given frequency range without considering the distance to the jet origin.

Overestimation of a distance to a jet base by the core shift method may lead to biases in geometry interpretation. Here we speculate that the observed transition to a conical jet shape in M87 on a sub-parsec scales \citep{Hada13} may be explained by the geometrical reasons, although we do not exclude purely physical reasons on such a small scale. 
Suppose a jet boundary shape follows the relation $r=a(z-z_0)^{k}$, with a base $z_0$ shifted in a jet direction. This may result from a discrepancy of a parabolic jet apex and the estimated jet base, from which the distances to radio cores are calculated.  Plotting the jet boundary shape in $r-z$ coordinates in logarithmic scale in this case will lead to effective downward bend of a straight line on scales of the order of $z_0$. Although the jet boundary does not follow the power law for $z\gtrsim z_0$, it may be approximated by some $r=az^k$ locally, with the power $k$ depending on a scale on which the approximation is performed. We tested this in application to M87, for which such a bend has been discovered by \citet{Hada13}. We find that setting a jet boundary as $r_{\rm pc}=0.07(z-z_0)^{0.56}$ with $z_0=0.00153$~pc (approximately 2$-$3 Schwarzschild radii $R_{\rm S}$) 
mimics the power law $r\propto z^{0.77}$ on scales of $10-30\;R_{\rm s}$, while following the parabolic trend $r\propto z^{0.56}$ further downstream \citep{Hada13}.
It is $r\propto z^{0.77}$ that must be used on the scales of the order of $z_0$ in the expression of the core shift exponent $k_z$ \autoref{k_distance_acc}. This position of an observed jet apex may point to purely Blandford--Znajek mechanism of a jet launch in M87. If the result is true, it may also provide information on the position of a surface where the plasma filling the jet is born. Here we should note that such scales are unattainable within the semi-analytical cylindrical approach, as the flow becomes essentially non-cylindrical at the critical surfaces \citep[see discussion in ][]{Beskin06}. So here we just assume that the outflow has a quasi-parabolic boundary on the scales of tens of gravitational radii. This boundary shape may be supported by the disc outflow as seen in numerical simulations by \citet{Nakamura+18}.
 
Setting $k_z = 4k/3$ for the accelerating jet in $\Gamma < 1/\theta$ region and $k=0.77$ we obtain $z\propto\nu^{-0.97}$, close to the measured by \citet{Hada11_M87} $z\propto\nu^{-0.94}$.

\section{Discussion} \label{RandD}

Our result is consistent with the jet shape measurements done by \citet{MOJAVE_XIV} using stacked VLBI images of the MOJAVE sample sources at 15 GHz. They found that all sources used in our analysis have quasi-parabolic jet shapes with $k < 1$ on milliarcsecond angular scales, except for BL\,Lac with $k = 1.11$. However, for this source the break of the jet shape was directly observed at $z_{\rm obs} = 2$ mas by \citep{Kovalev20_r1}. The authors also obtained $k$ measurements using stacked 15 GHz and single epoch 1.4~GHz VLBI data. For 3 sources in our sample (0415+379, 0430+052, 2200+420) the break was observed directly. For other 8 sources a single profile fitting resulted in 6 sources with significant $k < 1$. Thus, it is an additional evidence that those cores reside in the parabolic domain. 

Our results are consistent with findings of \cite{2016ApJ...826..135L} who compiled the core brightness temperatures at various radio frequencies (2, 8, 15 and 86 GHz) and found that it increases with increasing the distance from the central engine. They attributed this to the jet acceleration from $\Gamma \propto 1$ at the position of 86 GHz cores to $\Gamma \approx 40-50$ at the position of 8-15 GHz cores. These values are consistent with our estimates (see \autoref{sample_bes_par}). Although \cite{2016ApJ...826..135L} assumed $z_{\rm core} \propto \nu^{-1}$ while deriving the distances and, thus, ignored possible non-conical jet shape and bulk acceleration dependence of $k_z$ (see \autoref{s:coreshift}), this does not affect the $\Gamma$ values, but only the slope of the acceleration profile.

Our results imply that jets can accelerate up to relatively high Lorentz factors $\Gamma \sim 100$. However such big values do not contradict the statistics of the superluminal motion of the jet components in the MOJAVE sample \citep[see fig.7 in][]{kutkin19} from \citet{MOJAVE_XIII}. At the same time the velocity transverse stratification that is essential for MHD jet models implies that the fastest jet regions can be ``hidden'' from the observer either by the Doppler deboosting or by the non-uniform emissivity distribution across the jet \citep[e.g. see discussion in ][concerning statistical modelling of the parent population of the MOJAVE sample]{MOJAVE_XVII}. Indeed, high $\Gamma$ are consistent with the ``hollow'' jets observed at large viewing angles. {\it RadioAstron} results of 3C\,84 require fast spine with $\Gamma > 20$ to explain the spine -- sheath brightness ratio \citep{RA_3C84}. If the velocity stratification is the only reason for this effect, the spine has to be accelerated to $\Gamma > 10$ within the first few hundred $r_{\rm g}$. The similar situation holds for M87 where the ``hollow'' jet structure is seen already at 56 $r_{\rm g}$ at 86\,GHz \citep{2016ApJ...817..131H}. \citet{2016ApJ...833...56A} found that the spine with $\Gamma = 45$ viewed at 14$^\circ$ can explain the M87 jet inner ridgeline dimming at 5\,GHz. \citet{2018ApJ...855..128W} used stacked 43\,GHz VLBA images of M87 and estimated the minimal spine Lorentz factor $\Gamma = 16.5$ needed to explain the brightness asymmetry. They note that this is consistent with the apparent motion measurements in HST-1 in optical band~\citep{1999ApJ...520..621B} and fastest components seen in the radio~\citep{2012A&A...538L..10G}.

Our findings that acceleration continues on longer scales in high power jets is consistent with the findings of \citet{2015MNRAS.453.4070P}. They used an analytical jet model with a transition from a parabolic accelerating domain to a conical weakly decelerated region, fitted it to the spectral energy distribution of 37 blazars, and found that acceleration continues up to $z\approx200$\,pc for the most powerful sources.

Kinematics of jet components measured using multi-epoch VLBI images can reflect instability patterns with heated particles separated from the jet flow at different scales as suggested in K19 and observed in simulations of \citet{2021ApJ...907L..44S}. For a stratified outflow the kinematic measurements may also provide the real plasma speed but for a slower jet regions, e.g. a sheath flow velocity.  

The conclusion that the data is in agreement with the EA regime in a parabolic jet shape does not depend on whether we use M1 or M2 (Section~\ref{Sample}). However, the numerical estimates do depend on a particular model, with M2 providing systematically better agreement with the expected values of AGN physical properties (black hole spins) than M1. This means that a model with a sheath flow enclosing the fast spine may be preferential. In the M2 a slower sheath occupies about a half of the jet radius. We expect that the models with higher ratio of a sheath-to-spine width might describe AGN jet structure even better.

\section{Conclusions} \label{sec:concl}

Boundary shapes of AGN jets close enough and viewed at large angles are observed directly by the VLBI. Distant sources, however, are unattainable for such straightforward measurements due to the limited angular resolution, although accurate modeling of the core widths points to a parabolic geometry in some of them. In this work we propose a new method to probe the geometry of a jet at the region of its radio core. 
We apply the method to distant sources with small inclination angles, using the observational data at 8 and 15\,GHz by~\citet{kutkin19}. 

The results obtained for 11 sources indicate that radio cores are located in the efficient acceleration zone, and the jet boundary is quasi-parabolic. Within this scenario we can estimate BH spins using jet parabolic opening coeffients measured for the nearby sources. For four sources the estimate provides upper possible values $a_*\approx 0.1$, marginally consistent with the expected values. If the measured opening coefficients are upper limits due to limited resolution, the corresponding estimates will be a lower limit.
We demonstrate that if the cores are in effective acceleration parabolic regime, their position at the $\Gamma-z$ plane should occupy a more or less narrow band, which width depends weakly on the jet and ambient medium properties.
We create semi-analytical models of a jet transverse and longitudinal structure for the constant field lines angular velocity (M1), and for angular velocity dropping to zero at the jet boundary (M2).
By simulating either parabolic or conical boundary shape, we compare them with the observational data.
It is shown that sustaining effective acceleration regime for a conical jet requires unphysically large pressure at distances of order of 10\,pc downstream, and, therefore, this scenario can be rejected.
The scenario of cores being in the regime of acceleration saturation ($\Gamma\approx{\rm const}$) is also unlikely, because in this case there should be no relation between a Lorentz factor $\Gamma$ and a core position along a jet $z$.

We show that for a parabolic jet the distance between the base and the core is overestimated, if derived from the core shift within a conical geometry assumption. 
If the real apex in a quasi-parabolic jet is displaced downstream with respect to the estimated one, then the inferred geometry can be quasi-conical on scales of this displacement value. This explains both the quasi-conical jet shape observed on scales of 5$-$10 gravitational radii and the core-shift frequency dependence in the M\,87 jet.
We demonstrate that a jet geometry does not affect the core size frequency dependence $\theta_{\rm core} \propto \nu^{-1/k_{\theta}}$ directly with $k_{\theta}\approx1$. A less steep dependence, in turn, implies a bulk acceleration of the jet. 

The proposed method is robust since it relies on the universal bulk Lorentz factor behavior across a jet. In general, it requires well sampled light curves of a source flare at two frequencies and core shift measurements at the same frequencies to estimate the jet speed. The sample by \citet{kutkin19} was selected based on the light curves of the sources monitored at Michigan University. There are several other multifrequency single-dish monitoring programs\footnote{See the list at \url{http://www.physics.purdue.edu/MOJAVE/blazarprogramlist.html}}, e.g. F-Gamma project at Effelsberg telescope~\citep{2019A&A...626A..60A}, whose data can be used to increase the sample. This will be a promising project for future work.

\section*{Acknowledgements}

We thank the referee for the thoughtful and valuable comments that helped to improve this paper. We thank Evgeniya Kravchenko for valuable comments and suggestions. We thank Alexander Plavin for sharing the photosphere apparent speed estimates and their discussion. This study has been supported
by the Russian Science Foundation: project 20-62-46021.

\section*{Data availability}
The data underlying this article is available in \citet{kutkin19}.
This research made use of the data from the MOJAVE database\footnote{\url{http://www.physics.purdue.edu/MOJAVE/}} which is maintained by the MOJAVE team \citep{MOJAVE_XV}.
This research made use of NASA's Astrophysics Data System. This research has made use of the VizieR catalogue access tool, CDS,
Strasbourg, France (DOI : 10.26093/cds/vizier). The original description of the VizieR service was published in \citet{Vizier2000}.
The data generated
in this research will be shared on reasonable request to the
corresponding author.

\bibliographystyle{mnras}
\bibliography{nee}

\appendix

\section{Maximum Lorentz factor and magnetization parameter}
\label{a:sigma}

Usually we define the Michel's magnetization parameter
as
\begin{equation}
\sigma_{\rm M}=\frac{\Omega_0^2\Psi_0}{8\pi^2\mu\eta c^2},
\end{equation}
where $\Omega_0$ is a characteristic magentic field line angular velocity,
$\Psi_0$ is a total flux contained in a jet. For the a relativistic enthalpy 
$\mu=mc^2+mw$ we, as a rule, can neglect the non-relativistic enthalpy term.
The integral $\eta$ is defined so as
\begin{equation}
n{\textbf u}=\eta{\textbf B},
\label{eta}
\end{equation}
where a particle number density $n$ is defined in plasma proper frame and
$\textbf u$ is a plasma bulk four-velocity vector. This definition reflects
correctly the maximum Lorentz factor $\Gamma_{\rm max}$ that can be attained by plasma if all the
electromagnetic energy is transformed into plasma bulk motion kinetic energy within 
the jet model by Lyubarsky (2009).
However, for models like by Beskin et al. (2017) the value of $\sigma_{\rm M}$
is far off the value of $\Gamma_{\rm max}$. Let us relate the definition of the initial magnetization
$\sigma_{\rm M}$ with $\Gamma_{\rm max}$ for M1 and M2.

The magnetization parameter $\sigma$ is a ratio of a Poynting flux ${\textbf S}$
to particle energy flux ${\textbf K}$
\begin{equation}
\sigma=\frac{\left|{\textbf S}\right|}{\left|{\textbf K}\right|}.
\end{equation} 
The Poynting flux can be rewritten as
\begin{equation}
\left|{\textbf S}\right|=\frac{c}{4\pi}\left|{\textbf E}\times{\textbf B}\right|=\frac{\Omega_{\rm F}I}{2\pi c}
\left|{\textbf B}_{\rm P}\right|.
\end{equation}
Plasma energy flux
\begin{equation}
{\textbf K}=mc^2\Gamma n_{\rm lab} {\textbf v}_{\rm P}=mc^3\eta\Gamma{\textbf B}_{\rm P}.
\end{equation}
Here we use $n=n_{\rm lab}/\Gamma$ and the definition (\ref{eta}). Thus,
magnetization
\begin{equation}
\sigma=\frac{\Omega_{\rm F}I}{2\pi mc^4\eta\Gamma}.\label{sigma}
\end{equation}

Now we use this expression to define $\Gamma_{\rm max}$.
The integral of energy flux is defined as
\begin{equation}
E=\frac{\Omega_{\rm F}I}{2\pi c^2}+\mu\eta\Gamma.
\end{equation}
Dividing this by $\mu\eta\Gamma$ we get
\begin{equation}
\frac{E}{\mu\eta\Gamma}=\sigma+1.
\end{equation}
Now let us apply this to the particular choice of integrals
for different models. The model with a constant $\Omega_{\rm F}$ \citep{Beskin06, Lyu09} gives $\Omega_{\rm F}=\Omega_0$ and 
$$
E(\Psi)=\frac{\Omega_0^2\Psi}{4\pi^2 c^2}+\mu\eta\Gamma_{\rm in}.
$$
In this case,
$$
\Gamma\left(\sigma+1\right)=\frac{E}{\mu\eta}=\frac{\Omega_0^2\Psi}{4\pi^2 c^2\mu\eta}+\Gamma_{\rm in}=
2\sigma_{\rm M}\frac{\Psi}{\Psi_0}+\Gamma_{\rm in}.
$$
This is an important relation between $\Gamma$ and $\sigma$ in a jet, which depends on 
a field line. The maximum Lorentz factor can be achieved if $\Psi=\Psi_0$ and $\sigma\rightarrow\infty$.
In this case
\begin{equation}
\Gamma_{\rm max}=2\sigma_{\rm M}+\Gamma_{\rm in}.\label{sigma_lyu}
\end{equation}

In the jet model with a closed electric current inside a jet \citep{BCKN-17} the energy integral
is set to be
\begin{equation}
E(\Psi)=\frac{\Omega_0^2}{4\pi^2c^2}\Psi\left(1-\frac{\Psi}{\Psi_0}\right)+mc^2\eta\left[\Gamma_{\rm in}+\frac{\Psi}{\Psi_0}\left(1-\Gamma_{\rm in}\right)\right].
\end{equation}
Thus,
\begin{equation}
\Gamma(\sigma+1)=f(\psi),
\end{equation}
where $\psi=\Psi/\Psi_0$ and
$$
f(\psi)=2\sigma_{\rm M}\psi\left(1-\psi\right)+\Gamma_{\rm in}+\psi\left(1-\Gamma_{\rm in}\right).
$$
Function $f$ has a maximum at
$$
\psi_{\rm max}=\frac{2\sigma_{\rm M}+1-\Gamma_{\rm in}}{4\sigma_{\rm M}}\approx\frac{1}{2},
$$
and its value
\begin{equation}
f(\psi_{\rm max})=\frac{\sigma_{\rm M}+1+\Gamma_{\rm in}}{2}+\frac{\left(1-\Gamma_{\rm in}\right)^2}{8\sigma_{\rm M}}.
\end{equation}
The Lorentz factor is maximal when $\psi=\psi_{\rm max}$ and $\sigma=0$, so
\begin{equation}
\Gamma_{\rm max}=\frac{\sigma_{\rm M}+1+\Gamma_{\rm in}}{2}+\frac{\left(1-\Gamma_{\rm in}\right)^2}{8\sigma_{\rm M}}.\label{sigma_bes}
\end{equation} 

Comparing the expressions (\ref{sigma_lyu}) and (\ref{sigma_bes}) we see, that for the same
values of $\sigma_{\rm M}$, tha maximum Lorentz factor in a model by \citet{BCKN-17}
is roughly four times smaller than in the model by \citet{Lyu09}. Inversely, modelling the same
maximum Lorentz factor, we have to take $\sigma_{\rm M}$ four times larger for the model by \citet{BCKN-17}.

\section{Non-constant flow Lorentz factor}
\label{a:velocity}

Let the maximum velocity of a jet at given distance $z$ from the jet base be a function of $z$: $\beta(z)c$. For simplicity (in order to integrate analytically) we take $\Gamma(z)=a\sqrt{z}$, which leads to
\begin{equation}
\beta(z)=\sqrt{\frac{a^2z-1}{a^2z}}.
\end{equation}
We rewrite the equation (2) from 
K19 in the following form:
\begin{equation}
\frac{d z}{cd t_{\rm obs}}=\frac{\beta(z)}{1-\beta(z)\cos\theta}.
\end{equation}
The integration from $z_{15}$ to $z_8$ gives
\begin{equation}
I(z_{15},\;z_{8})-\Delta z\cos\theta=c\Delta t_{\rm obs}, 
\label{betanonconst}
\end{equation}
where
\begin{equation}
I=\left.\frac{1}{a^2}\left(\sqrt{a^2z(a^2z-1)}-\frac{1}{2}\ln\frac{\sqrt{a^2z}-\sqrt{a^2z-1}}{\sqrt{a^2z}+\sqrt{a^2z-1}}\right)\right|_{z_{15}}^{z_8}.
\end{equation}
The measurements by K19 assumed the constant velocity $\beta_{\rm const}$. In this case the equality
\begin{equation}
\frac{\Delta z}{c\Delta t_{\rm obs}}=\frac{\beta_{\rm const}}{1-\beta_{\rm const}\cos\theta}
\label{betaconst}
\end{equation}
holds with $\beta_{\rm const}$ --- the constant flow velocity obtained by K19. Equating \autoref{betanonconst} and \autoref{betaconst}
we obtain
\begin{equation}
\beta_{\rm const}=\frac{\Delta z}{I(z_{15},\;z_{8})}.
\label{bI}
\end{equation}
To understand how far off our results for which we use assumption of a constant $\Gamma$, we compare the Lorentz factor at $z_{\rm core}$ within the model of accelerating flow with the constant Lorentz factor we use.  To do this, we choose the coefficient $a\approx 2.7$ for each of the sources so as to have \autoref{bI}. After this, using $a$, we calculate the Lorentz factor $\Gamma_{\rm var}$ at $z=z_{\rm core}=(z_8+z_{15})/2$. The obtained $\Gamma_{\rm var}$ are 1.5$-$2\% larger than the observed $\Gamma$.
Thus, our use the Lorentz factors, obtained under assumption of a constant flow velocity, is justified for the effectively accelerating flow. 

\section{Sources with previously detected parabolic jet boundary} \label{Three}

Our sample of 11 sources include 3C 111, 3C 120 and BL Lac with the earlier detected jet boundary shape break \citet{Kovalev20_r1}. This detection was possible due to the sources proximity (redshift $<0.07$), which provided the necessary linear resolution. Our aim is to check the consistency of the analytical and semi-analytical modelling for these sources, data on the value of a Lorentz factor and the additional data on the parabolic jet shape. For these three sources we calculate the dependence $\Gamma(r)$ within semi-analytical approach \citep{BCKN-17} for two models: with the constant magnetic field lines angular velocity (M1) and with an electric current, closed inside a jet (M2). In order to obtain the dependence $\Gamma(z)$ we use the fitted parabolic jet boundary shapes from \citet{Kovalev20_r1}, extrapolating it inside down to the core scales. We have one free parameter to fit the data point: maximum Lorentz factor $\Gamma_{\rm max}$. Setting it to a certain value we can find the corresponding light cylinder radius $R_{\rm L}$. The last one can be related to the black hole spin if we assume the maximum jet power output condition $\Omega_{\rm F}/\Omega_{\rm H}=1/2$ \citep{BZ-77}. The values of observational angles given by K19 and by \citet{Kovalev20_r1} are different, and we employ the first value for the uniformity of the sample. We also use the black hole mass for a particular source collected in \citet{Kovalev20_r1}.

\begin{figure}
\centering
\includegraphics[width=\columnwidth, trim=0cm 1cm 0cm 0cm]{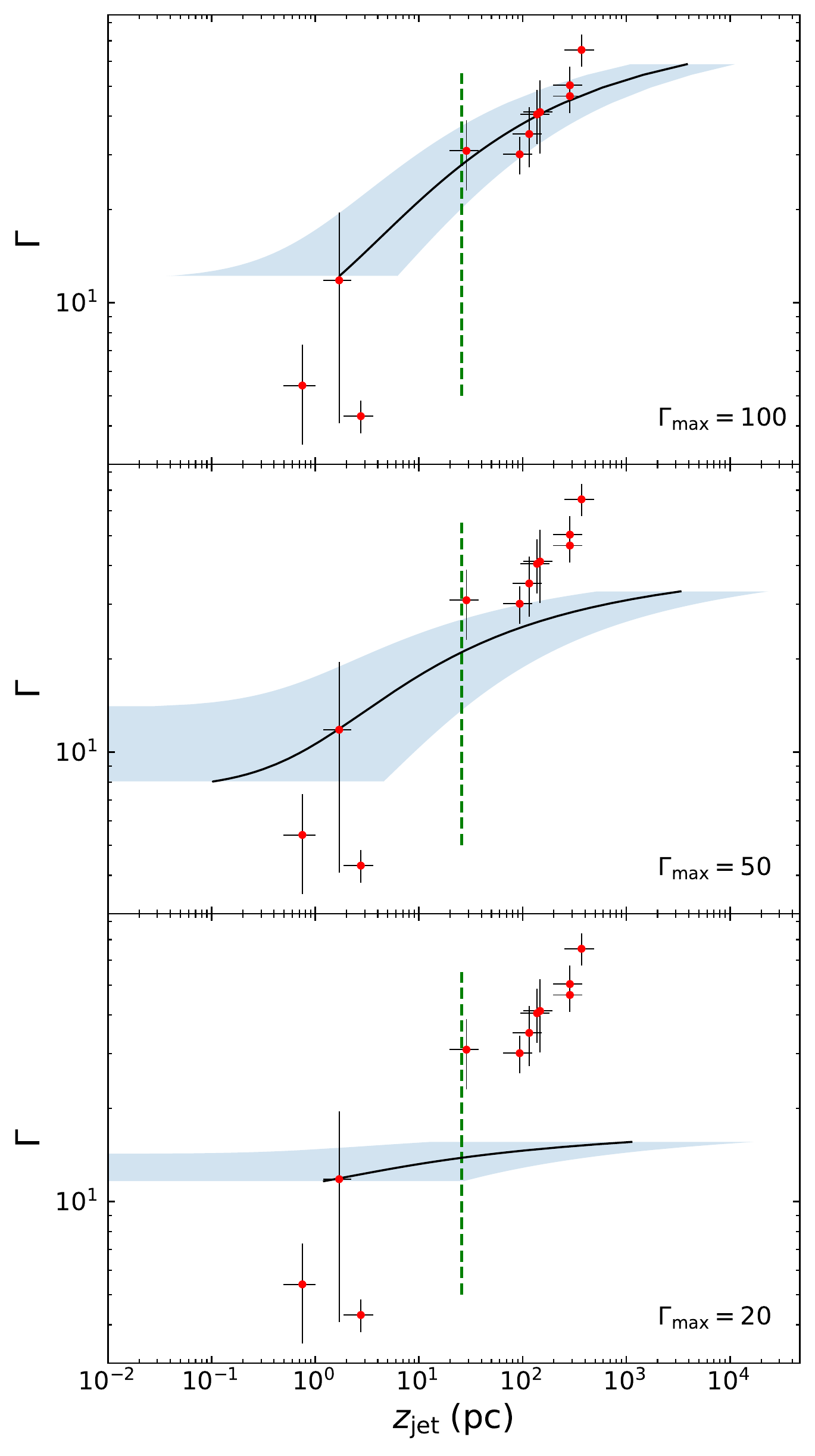}
\caption{Lorentz factor fits for BL Lac. Red dots are data point with errors in black thin lines. Black curves represent the modelled dependence within the M2, grey shaded region enclose the error in Lorentz factor (lower limits are approximate, upper limit for $\Gamma_{\rm max}=20$ is less than error bar ending due to initial magnetization choice). Green dashed line designates the detected by \citep{Kovalev20_r1} jet shape break. The curve, corresponding to $\Gamma_{\rm max}=100$, is fitted setting $a_*=0.014^{+0.014}_{-0.006}$. $\Gamma_{\rm max}=50$ provides $a_*=0.019^{+0.031}_{-0.013}$, and $\Gamma_{\rm max}=20$ provides $a_*=0.093^{+0.707}_{-0.071}$.}
\label{2200lyu}
\end{figure}

For BL Lac we fit the data for the given maximum Lorentz factor $\Gamma_{\rm max}=20,\;50,\;100$. Note here, that in this work we use different definition of $\Gamma_{\rm max}$ in comparison with \citet{Kovalev20_r1,NKP20_r2} (see Appendix~\ref{a:sigma}). To fit the data we set a jet boundary shape exactly as $r=0.505(z+0.087)^{0.537}$ \citep{Kovalev20_r1}. Then we find such an $R_{\rm L}$ that the semi-analytical curve $\Gamma(z)$ fits the data point within the error bars (see \autoref{2200lyu}). 
For the M2 we are able to fit the data (see \autoref{2200lyu})
setting $R_{\rm L}=86,\;431,\;5540\;r_{\rm g}$ (we assume $M=10^{8.23}$ \citep{WU02}),
corresponding to the black hole (BH) spin values $a_*=0.093^{+0.707}_{-0.071},\;0.019^{+0.031}_{-0.013},\;0.014^{+0.014}_{-0.006}$. 
We note that fitting the kinematics provides black hole spin much larger than the value obtained by fitting the jet shape boundary break \citep{NKP20_r2}. The green dashed line in \autoref{2200lyu} designates the jet break obtained by \citep{Kovalev20_r1}. For the lowest semi-analytical curve corresponding to $\Gamma_{\rm max}=20$ the transition from effective acceleration to saturation is hardly distinguished in the plot in the scale. We also note that the furthest cores do not lie on the bending part of the curve. They have much faster dependence of $\Gamma$ of $z$.

We are able to fit the data for 3C 111 and 3C 120 with $\Gamma_{\rm max}=10,\;20$. The models with higher $\Gamma_{\rm max}$ start further downstream, so applying these models would require extrapolating the analytical curves upstream. However, we observe, that fitting the particular point in $z$---$\Gamma$ plane provides smaller spin for larger $\Gamma_{\rm max}$ (unlike fitting the jet shape break in \citet{NKP20_r2}), so taking smaller $\Gamma_{\rm max}$ we model larger BH spin values. This is easy to understand. We try to fit the observational point $\{\Gamma_*,\;z_*\}$ in $\Gamma-z$ coordinate plane. The universal power law for a chosen source is given by $\Gamma=\alpha_1\sqrt{\theta z}/2R_{\rm L}$. These curves for different $\Gamma_{\rm max}$ have one envelope and do not intersect. In order to draw them through one  point we need to choose larger $R_{\rm L}$ for larger $\Gamma_{\rm max}$. This means that larger $\Gamma_{\rm max}$ correspond to smaller $a_*$. For 3C 111 we obtaing for M2, $\Gamma_{\rm max}=10$ $a_*=0.019^{+0.009}_{-0.005}$. For $\Gamma_{\rm max}=20$ $a_*=0.012^{+0.002}_{-0.002}$. 
The 3C 120 black hole spin values for M2, $\Gamma_{\rm max}=10$ are $a_*=0.052^{+0.432}_{-0.037}$. For $\Gamma_{\rm max}=20$ $a_*=0.021^{+0.019}_{-0.008}$.

Here we confirmed that the observed cores at $8$ and $15$~GHz are in EA parabolic region for these sources.
We have estimated black hole spins for these sources, and they have systematically higher values in comparison with the results based on fitting the jet shape break \citep{NKP20_r2}. However, the result depends strongly on a particular choice of $\Gamma_{\rm max}$.

\bsp    
\label{lastpage}
\end{document}